\documentclass[fleqn,usenatbib]{mnras}

\usepackage{newtxtext,newtxmath}

\usepackage[T1]{fontenc}

\DeclareRobustCommand{\VAN}[3]{#2}
\let\VANthebibliography\thebibliography
\def\thebibliography{\DeclareRobustCommand{\VAN}[3]{##3}\VANthebibliography}

\usepackage{graphicx}	
\usepackage{amsmath}	
\usepackage{amssymb}	

\newcommand{\minew}[1]{{\color{black}{#1}}}
\newcommand{\miold}[1]{\iffalse{#1}\fi}

\title[Transitional dark energy]{Can phantom transition at $z\sim 1$ restore the Cosmic concordance?}

\author[Zhihuan Zhou]{
Zhihuan Zhou,$^{1}$\thanks{E-mail: 11702005@mail.dlut.edu.cn}
Gang Liu,
Yuhao Mu,
Lixin Xu\thanks{E-mail: lxxu@dlut.edu.cn}
\\
Institute of Theoretical Physics, School of Physics, Dalian University of Technology, Dalian, 116024, China
}

\date{Accepted 2022 January 4. Received 2021 December 21; in original form 2021 October 21}

\pubyear{2022}

\begin{document}
\label{firstpage}
\pagerange{\pageref{firstpage}--\pageref{lastpage}}
\maketitle

\begin{abstract}
The tension among inferences of Hubble constant ($H_0$) is found in a large array of datasets combinations.
Modification to the late expansion history is the most direct solution to this discrepancy. 
In this work, we examine the viability of restoring the cosmological concordance with a novel version of transitional dark energy (TDE). The main anchors for the cosmic distance scale: cosmic microwave background (CMB) radiation, 
baryon acoustic oscillation (BAO), and Type Ia supernova (SNe Ia) calibrated by Cepheids form a ``impossible trinity'', 
i.e., it's plausible to reconcile with any two of the three but unlikely to accommodate them all. Particularly, 
the tension between BAO and the calibrated SNe Ia can not be reconciled within the scenarios of late dark energy.
Nevertheless, our analysis suggests that the TDE model can reconcile with CMB and SNe Ia calibrated by its absolute magnitude ($M_{\rm{B}}$). 
when the equation of state (EoS) of DE transits around $z\sim1$. Meanwhile, we see a positive sign that the EoS transits with the \minew{inclusion of a local prior on $M_{\rm{B}}$, whereas the opposite is true without the $M_{\rm{B}}$ prior.} 
\end{abstract}




\section{Introduction}
The precise determination of the Hubble constant $H_0$, which requires both accurate astrophysical and 
cosmological modeling, is crucial for modern cosmology. 
Historically, $H_0$ is measured directly via the classical distance ladder (\citet{Sandage:2006cv}) in 
the local universe, while in the early universe, $H_0$ is extracted indirectly from angular size of 
the sound horizon $r_s$  and the shapes of the acoustic peaks in the cosmic microwave background (CMB) radiation (\citet{Hinshaw_2013,Ade:2013zuv}). These measurements disagree at $4\sim6\sigma$ significance, 
e.g., $H_0=73.2 \pm 1.3$ from SH0ES collaboration (\citet{Riess:2020fzl}) and $H_0=67.4\pm0.5$ from 
\emph{Planck} 2018 data\footnote{The obtained $H_0$ by alternative methods varies between these measurements. One example is the recent measurement $H_0 = 69.8\pm2.2$ from tip of the red giant branch (TRGB)
	calibration (\citet{Freedman:2021ahq}.} (\citet{Aghanim:2018eyx}). 
Several independent $H_0$ probes of the late-universe have reached competitive accuracy with the 
distance ladder, such as measurements of the strongly lensed quasar systems (\citet{Wong:2019kwg}), 
obtaining $H_0 =73.3_{-1.8}^{+1.7}$. Since the $H_0$ tension between late and early 
universe is found in a large array of differing datasets combinations (see \citet{Verde:2019ivm, Perivolaropoulos:2021jda} 
for a recent review), it's unnatural to attribute the disagreement solely to systematic errors\footnote{
	Extensive works have focused on whether this discrepancy can
	be caused by yet unknown systematic effects. Recent work in \citet{Mortsell:2021nzg} suggest that, by choosing different Cepheid colorluminosity calibration method, the obtained $H_0$ could be much more compatible with the Planck value.
	To show the application of the late dark energy to the $H_0$ fit, we will focus on the SH0ES result
	that has the largest deviation from the Planck result.}. 
Hence, this tension may be a sign of physics beyond the standard $\Lambda$CDM cosmology. 

There exist varied theoretical proposals to explain or ameliorate the $H_0$ discrepancy, 
ranging from new physics in early-time ($z\gtrsim 1100$, pre-recombination) to 
late-time ($z\lesssim 2$) universe (see \citet{DiValentino:2021izs} for a review).
Within the framework of General Relativity (GR), it is natural to consider the 
dynamical DE, i.e., replacing cosmological constant $\Lambda$ with a scalar field
(\citet{Barboza:2008rh,ArmendarizPicon:2000dh,Linder:2003,Jassal:2004ej,ArmendarizPicon:2000ah,Caldwell:2002,Guo:2005,DAmico:2020kxu}) or a DE equation of state (EoS) parameterization (see \citet{Li_2011} for a review). Whereas, it is proved the converse that Quintessence (a minimally coupled scalar) always makes Hubble tension worse than $\Lambda$ (\citet{Banerjee:2020xcn}).
Other tentative solutions include interacting DE 
(\citet{Kumar:2016zpg,Yang:2018euj,Pan:2020zza,DiValentino:2019ffd,Nunes:2021zzi}), 
modified gravity (\citet{CANTATA:2021ktz,Nunes:2018xbm,Raveri:2019mxg,Yan:2019gbw,Frusciante:2019puu,DAgostino:2020dhv,Ballardini:2020iws}) 
as well as other possibilities 
(\citet{Hart:2017ndk,Chiang:2018xpn,Yang:2019jwn,Jedamzik:2020krr,Sekiguchi:2020teg,Vagnozzi:2019ezj}).
Recently, the binned analysis of $H_0$ is employed in (\citet{Wong:2019kwg,Dainotti:2021pqg}) to investigate a possible evolutionary trend for $H_0$.

These proposals must also stand up to the scrutiny of current Large-Scale Structure (LSS) surveys 
(\citet{Abbott:2017wau,Alam:2016hwk,Hildebrandt:2018yau,Hikage:2018qbn}), 
which have delivered precise cosmological constraints.  
Interestingly, the inference of $H_0$ value in both 
CMB and LSS surveys depend on the precise determination of sound horizon $r_s$, 
thus, a reduction of the sound horizon by increasing the expansion rate just prior to recombination 
(with additional energy density components, e.g.,
exotic early dark energy (EDE) (\citet{Poulin:2018cxd,Das:2020wfe}), acoustic dark energy (ADE) (\citet{Lin:2020jcb}) 
seems to be the ``least unlikely to be successful'' (\citet{Knox:2019rjx}) approach to solve the $H_0$ tension.  
However, the addition of EDE or another component would suppress the growth of perturbations prior to 
recombination, which changes the amplitudes and phases of the CMB acoustic peaks in complex ways 
and would bring in new tension with the density fluctuation amplitude, $\sigma_{8}$.
Moreover, given a certain shift in $r_s$, the shift in $H_0$
needed in the CMB to match observations will be different from the one needed to match
the LSS observations, introducing a currently non-existing tension (\citet{Hill:2020osr}). 

In the late-DE scenarios, the sound horizon at last scattering $r_s(z_*)$ 
is preserved from the modification of late expansion history. 
With the acoustic scale ($D_A(z_*)/r_s(z_*)$) fixed by CMB, 
an upward shift on $H_0$ is compensated by the increase
of $D_A(z_*)$, which can be implemented by raising the ``dilution rate'' of DE density.
This scenario can be realized by a wide class of ``phantom-like'' 
DE (\citet{Caldwell:2002,Caldwell:2003vq,Guo:2005,Alestas:2020mvb}). 
\minew{Although the phantom transition of DE at late time is able to provide an apparent resolution of the $H_0$ tension, the underlying tension of the SnIa absolute magnitude ($M_{\rm{B}}$) have not been taken into account.
Recent works in \citet{Camarena:2021jlr,Benevento:2020fev,Alestas:2020zol} have studied phantom transition at very-low redshift $(z<0.1)$ accompanied by an $M_B$ transition or a rapid transition of the effective gravitational constant ($G_{\rm{eff}}$) at $z<0.01$ (\citet{Marra:2021fvf}), in order to address both the $H_0$ tension and $M_B$ tension. Being backed up by several observational hints (\citet{Alestas:2021nmi,Perivolaropoulos:2021bds}), this double transition model 
proves to be successful in addressing the $S_8$ tension as well (\citet{Marra:2021fvf}).}
In the recently proposed phenomenological emergent dark energy model (PEDE) (\citet{Li:2019yem}) model and its
generalized version (\citet{Li:2020ybr,Hernandez-Almada:2020uyr}), the DE is negligible at early times but dominates at late-time, providing an alternative solution to the coincidence problem (\citet{Hernandez-Almada:2020uyr}. 
If only the local $H_0$ measurements and CMB data are taken into consideration, 
this scenario seems ideal to solve the $H_0$ tension because the imprints of late-universe modification on the CMB spectra can be counteracted by a shift in $H_0$. However, the numerical analysis in (\citet{Rezaei:2020mrj,Benaoum:2020qsi}) suggests that, 
the PEDE model is not well compatible with SN Ia and LSS surveys, especially the $f\sigma_8(z)$ data from RSD observations (\citet{Rezaei:2020mrj}). Such incompatibility is mainly caused by the sharp transition of EoS at present, hence one may consider altering the scale at which the EoS of DE transits to avoid the tension.
Using Gaussian process regression, \citet{Keeley:2019esp} suggests that when the EoS transits rapidly at $1 \lesssim z \lesssim 2$, the growth of the perturbation is slower than fiducial $\Lambda$CDM model for $z\lesssim 1$, while the study in  \citet{Alestas:2021xes} suggest that deforming $H(z)$ with CPL parameterization worsens the growth tension.
As a result of these concerns, we proposed a novel version of transitional dark energy 
(TDE) model (based on the research in \citet{Keeley:2019esp,Bassett:2002qu}), which is captured by the transitional scale $a_c$. In this scenario, the DE behaves like the cosmological constant $\Lambda$ at late-time and goes through a rapid transition in the EoS at $a_c$.

The outline of this paper is as follows: in section \ref{sec:ph} we take a brief review of the key observations associated with $H_0$ tension. 
In section \ref{sec:models}, we briefly introduce the TDE model as well as its
cosmological features. In section \ref{sec:data}, we describe our numerical implementation of EDE model and the datasets used in our analysis. 
The numerical results of the Markov chain Monte Carlo (MCMC) analysis are presented in
section \ref{sec:results}. The discussion and conclusions are presented in section \ref{sec:conclusion}.
\section{MATCHING THE CMB}\label{sec:ph}
Measurements of the CMB spectra precisely determine the angular acoustic scale $\theta_s$ 
(transverse direction) and the ``shift parameter'' R (line-of-sight direction), defined as (\citet{Chen:2018dbv})
\begin{eqnarray}
\theta_s =r_s(z_*)/D_A(z_*), \qquad R(z_*)\propto \omega_{m}^{\frac{1}{2}} D_A(z_*),\label{eq:horizon}
\end{eqnarray}
where $z_*$ is the redshift at the photon decoupling epoch, 
$D_A(z) = \int^{z}_{0}\frac{dz'}{H(z')}$ is the angular
diameter distance and 
$r_s$ is the comoving sound horizon, defined by
\begin{eqnarray}
r_s(z_*) = \int^{\infty}_{z_*} \frac{dz}{H(z)}c_s(z), \label{eq:rs}
\end{eqnarray}
where $c_s(z)$ is the sound speed of the coupled photon-baryon fluid. 
In matter-radiation dominated era, an approximation of Hubble rate reads
\begin{eqnarray}
H(z) \propto \left[ \omega_r (1+z)^4 + \omega_m(1+z)^3 \right]^{\frac{1}{2}},
\end{eqnarray}
which have no dependence on $H_0$ when $\omega_m \equiv \Omega_{m,0}h^2$ and $\omega_{r} \equiv \Omega_{r,0}h^2$ are fixed at given values (hereinbelow). Thus, a shift of the local expansion rate $H_0$ as well as modifications to the expansion history after last scattering ($z < z_*$) preserves the sound horizon $r_s(z_*)$.

\begin{figure}
	\includegraphics[scale=0.28]{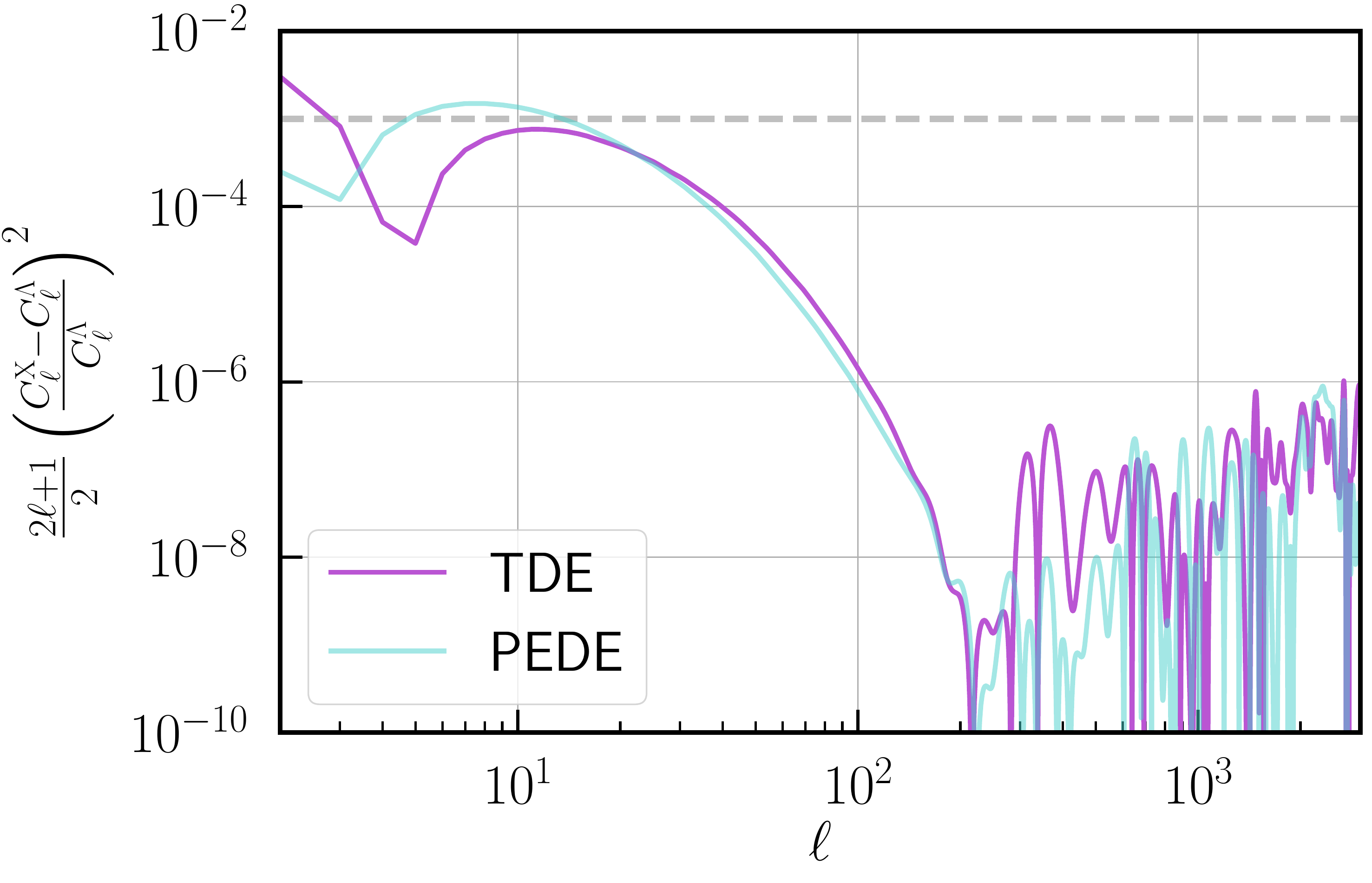}
	\caption{The disparity between the angular power spectrum calculated in  
		TDE, PEDE (\citet{Li:2019yem}) model ($C^{\rm{X}}_{\ell}$) and in $\Lambda$CDM model ($C^{\Lambda}_{\ell}$) divided by the cosmic variance, i.e., $\frac{2\ell+1}{2}\left(\frac{C^{\rm{X}}_{\ell}-C^{\rm{\Lambda}}_{\ell}}{C^{\rm{\Lambda}}_{\ell}}\right)^2$. Noting that $H_0$ is set to the best-fit value obtained by ``shooting'' (see section \ref{sec:ph}), while other parameters are set to the \emph{Planck} best-fit values.}\label{fig:perturb}
\end{figure}

Within the scenarios of late-DE, the DE component has no effective presence in the past, and thus have
no impact on $r_s(z_*)$.
Unlike EDE changes the power spectrum in a complex way (\citet{Poulin:2018dzj}), the presence of late-DE changes only the traveling distance of the free streaming photons in the late universe, resulting in a linear shift in power spectrum. Such a shift can be perfectly offset by adjustment of $H_0$ value, thus, the best-fit $H_0$ can be conveniently obtained by ``shooting''\footnote{By ``shooting'', we mean that when a default value of $D_A(z*)$ is chosen, we adjust $H_0$ until a suitable value is found which gives the correct $D_A(z*)$. This algorithm is frequently used in \texttt{CLASS} (\citet{Lesgourgues:2011re}.} the \emph{Planck} best-fit $D_A(z_*)$ in $\Lambda$CDM model\footnote{As is indicated in Eq. (\ref{eq:horizon}), the key features of the CMB spectrum ($\theta_s$ and $R(z*)$) depend only on  $D_A(z_*)$ when $\omega_m$ and $r_s$ are preserved.}. 
To verify the above argument, we calculate the disparity between the angular power spectrum calculated in 
$\Lambda$CDM model and in late-DE scenarios ($C^{\rm{\Lambda}}_{\ell}$), e.g., TDE, PEDE (\citet{Li:2019yem})
calibrated by the best-fit $H_0$. The results shown in Fig. \ref{fig:perturb} indicates that the differences is much smaller than the cosmic variance, i.e.,
\begin{eqnarray}
\frac{2\ell+1}{2}\left(\frac{C^{\rm{X}}_{\ell}-C^{\rm{\Lambda}}_{\ell}}{C^{\rm{\Lambda}}_{\ell}}\right)^2<0.01.
\end{eqnarray}
Consequently, the influence of late-DE on CMB is equivalent to a $H_0$ shift which preserves the overall shape of the power spectrum. On the other hand, for a given $H_0$ prior, the \emph{Planck}-$\Lambda$CDM best-fit spectrum can always be matched by adjusting late-DE parameters, thus, the tension between SH0ES and CMB is resolved within late-DE scenarios.
It should be noted that, in this work DE is treated as an effective fluid with equations of evolution for
density perturbation $\delta_{\rm{fld}}$ and velocity  $\theta_{\rm{fld}}$, respectively, given in (\citet{Ballesteros:2010ks}).
Solving these equations requires the specification of the fluid EoS $w(z)$ (see discussion in section \ref{sec:models}), the effective sound speed of DE $\hat{c}^2_s\equiv \delta p/\delta\rho$ (evaluated in the fluid rest frame), and adiabatic sound speed $c^2_a \equiv \dot{p}/\dot{\rho}$. We fix $\hat{c}^2_s=1$ so that the fluid can not be clustered, and $c^2_a = w - \dot{w}/3 \mathcal{H}(1+w)$  with $\mathcal{H} \equiv aH$ the
conformal Hubble rate.

\begin{figure}
	\includegraphics[scale=0.38]{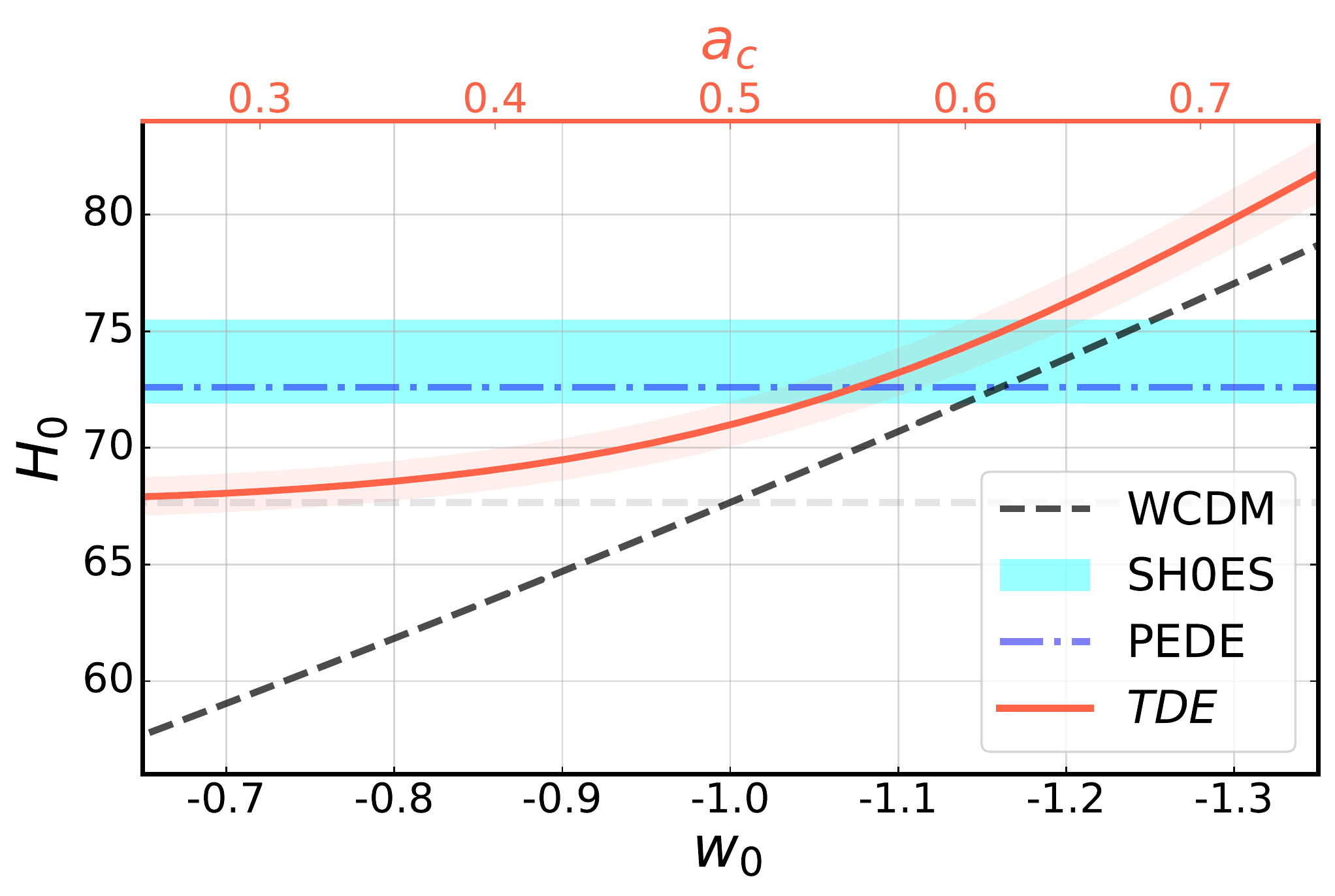}
	\includegraphics[scale=0.38]{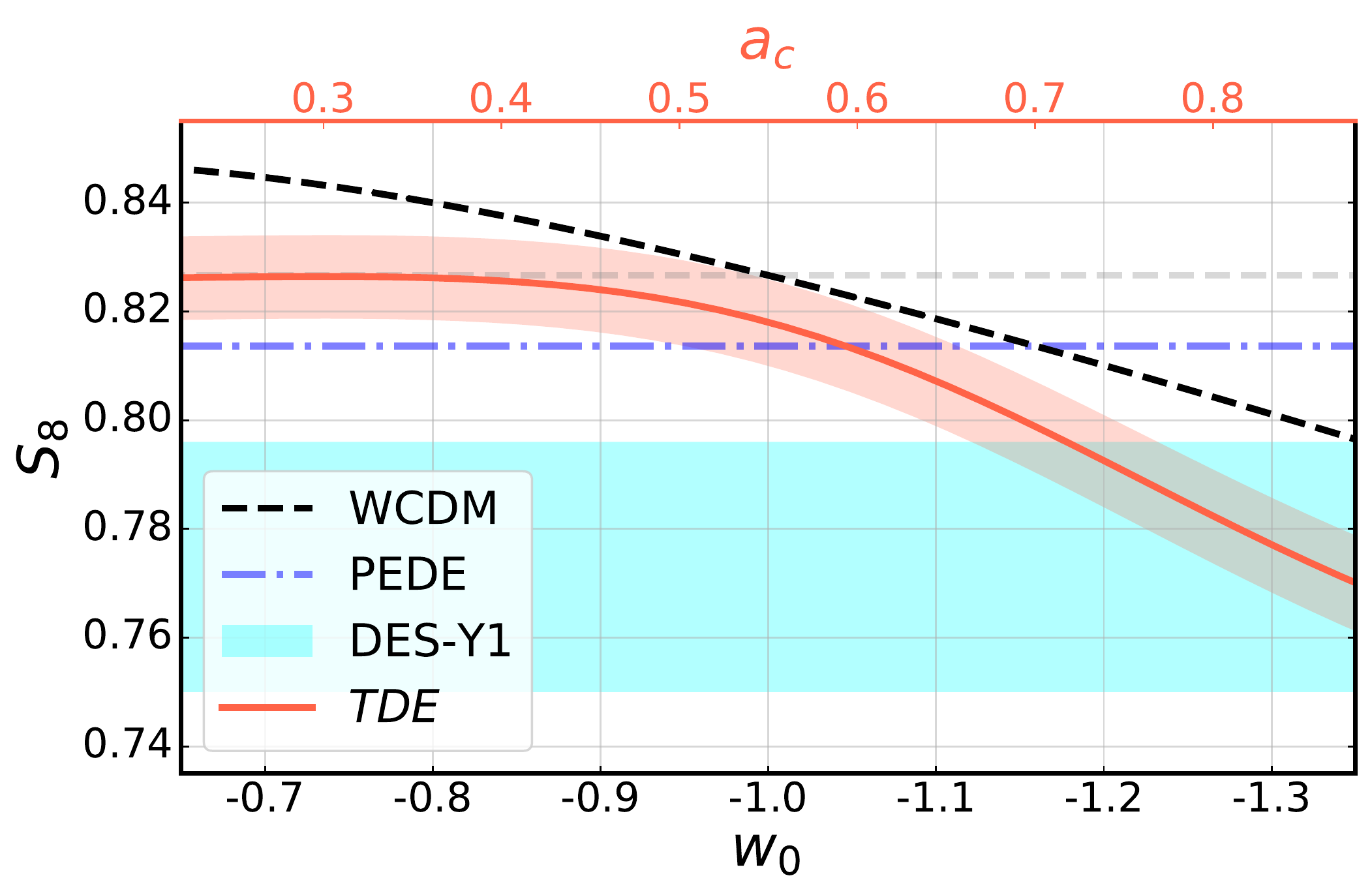}
	\includegraphics[scale=0.38]{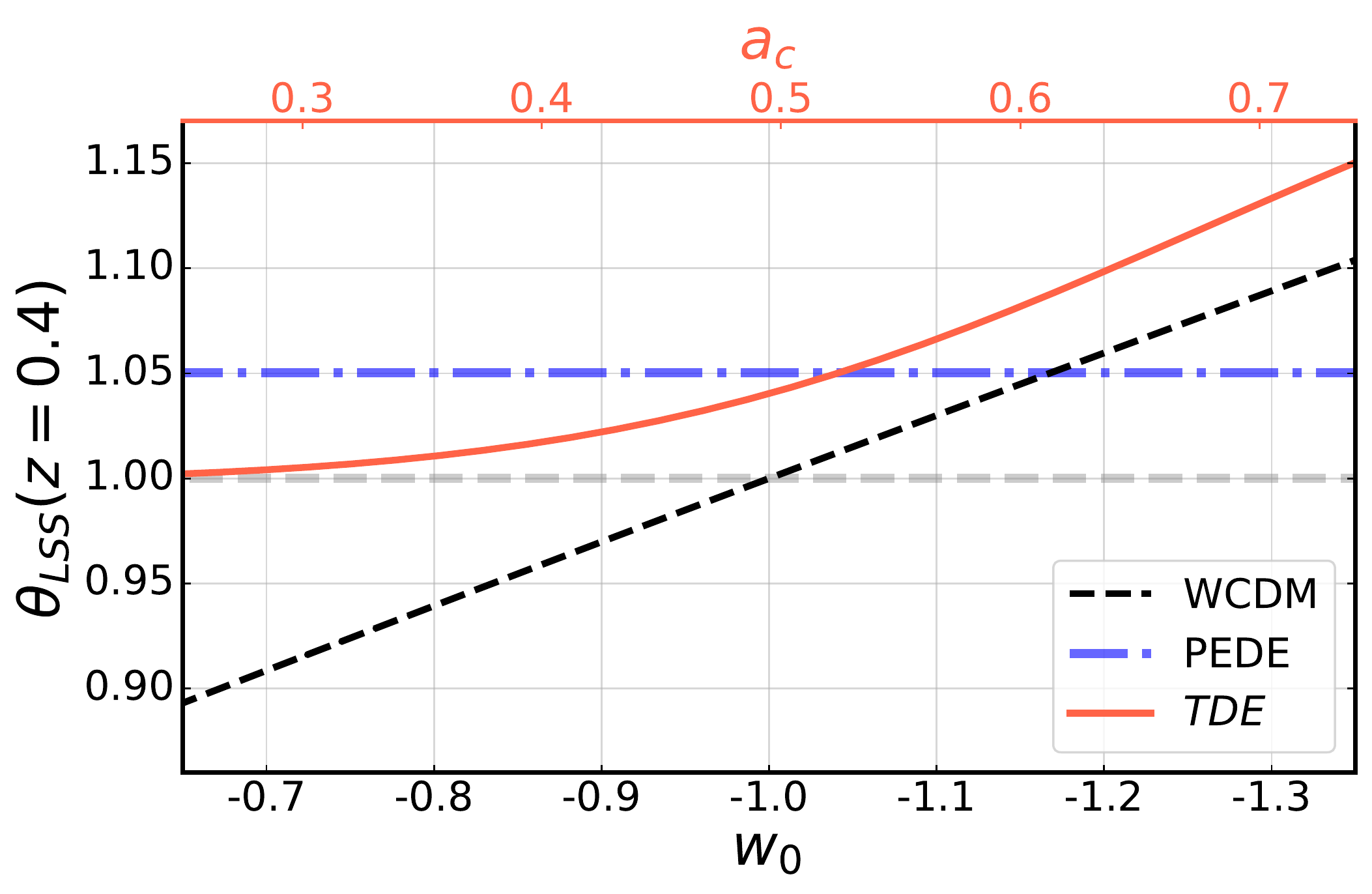}
	\caption{The \emph{Planck} best-fit $H_0$ (top panel), $S_8$ (middle panel) and $\theta_{\rm{LSS}}$ (bottom panel) for
		different late-DE models. 
		Noting that the best-fit observable are obtained by ``shooting'' $D_A(z_*)$ (see section \ref{sec:ph}). 
		The shaded regions correspond to the error propagation from $68\%$ confidence levels of the \emph{Planck} distance prior.
		The horizontal band correspond to the SH0ES constrains i.e  $H_0=73.2 \pm 1.3$ (top panel) and
		DES-Y1 constrains i.e $S_8 =0.773^{+0.026}_{-0.020}$ (middle panel).   
	}\label{fig:shooting_results}
\end{figure}

With \emph{Planck} the best-fit $H_0$ value, we can calculate other key observables ($S_8,f\sigma_8(z)$) to see whether the solution of $H_0$ tension is at the cost of bringing in new tensions with other observations.
For example, in LSS surveys the angular acoustic scale $\theta_s$ could also be extracted from galaxy power spectrum at low-redshft ($z\approx z_{\rm{LSS}}\sim 0.3$), denoting as:
\begin{eqnarray}
\theta_{\rm{LSS}} \simeq \frac{r_s(z_*)}{(D_A(z_{\rm{LSS}})^2\cdot cz_{LSS}/H(z_{\rm{LSS}}))^{\frac{1}{3}}},
\end{eqnarray}
with approximation $r_s(z_*)\sim r_s(z_d)$, where $z_d$ is redshift at drag epoch, and $D_{H}(z)\equiv1/H(z)$. 
We show in Fig. \ref{fig:shooting_results} the ``shooting'' results of the WCDM model (negative cosmological constant) (\citet{Visinelli:2019qqu}) and TDE model, the details of the TDE model will be discussed in the next section).

\begin{figure}
	\centering
	\includegraphics[scale=0.3]{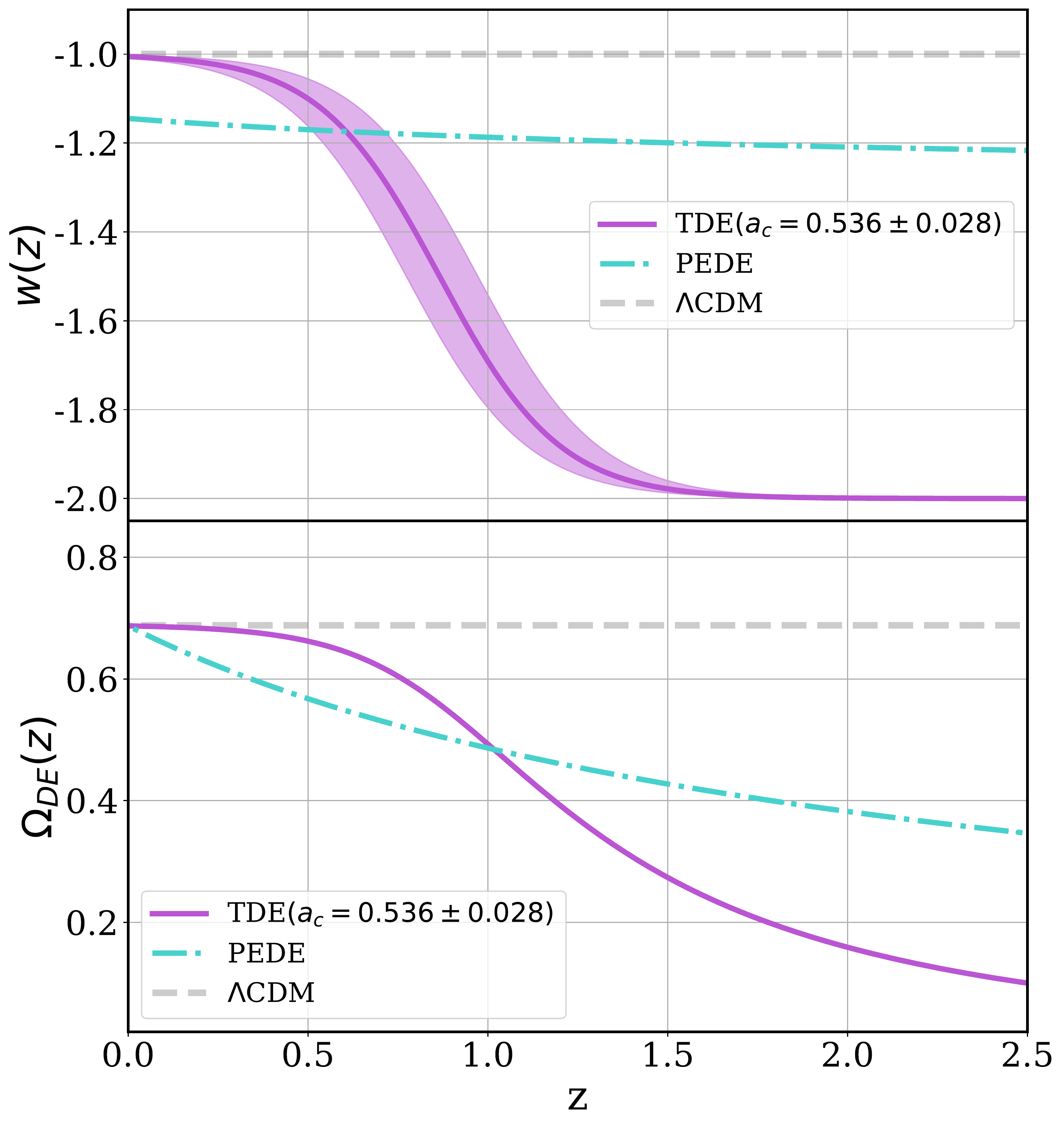}
	\caption{The evolution of EoS of DE (upper panel) and density ratio $\Omega_{\rm{DE}}\equiv\rho_{\rm{DE}}/\rho_{\rm{crit},0}$ (lower panel) for different DE models, respectively. The parameters are set in terms of the best-fit
		values of the joint \emph{Planck}+$M_{\rm{B}}$+SNe datasets, e.g., $a_c=0.536\pm0.028$ (see Table \ref{tab:accumulate_constraints}). 
		The shaded bands correspond to the 68\% confidence levels, while the gray dashed line corresponds to the $\Lambda$CDM model as a references.
	}\label{fig:w_fld}
\end{figure}

\begin{figure}
	\includegraphics[scale=0.3]{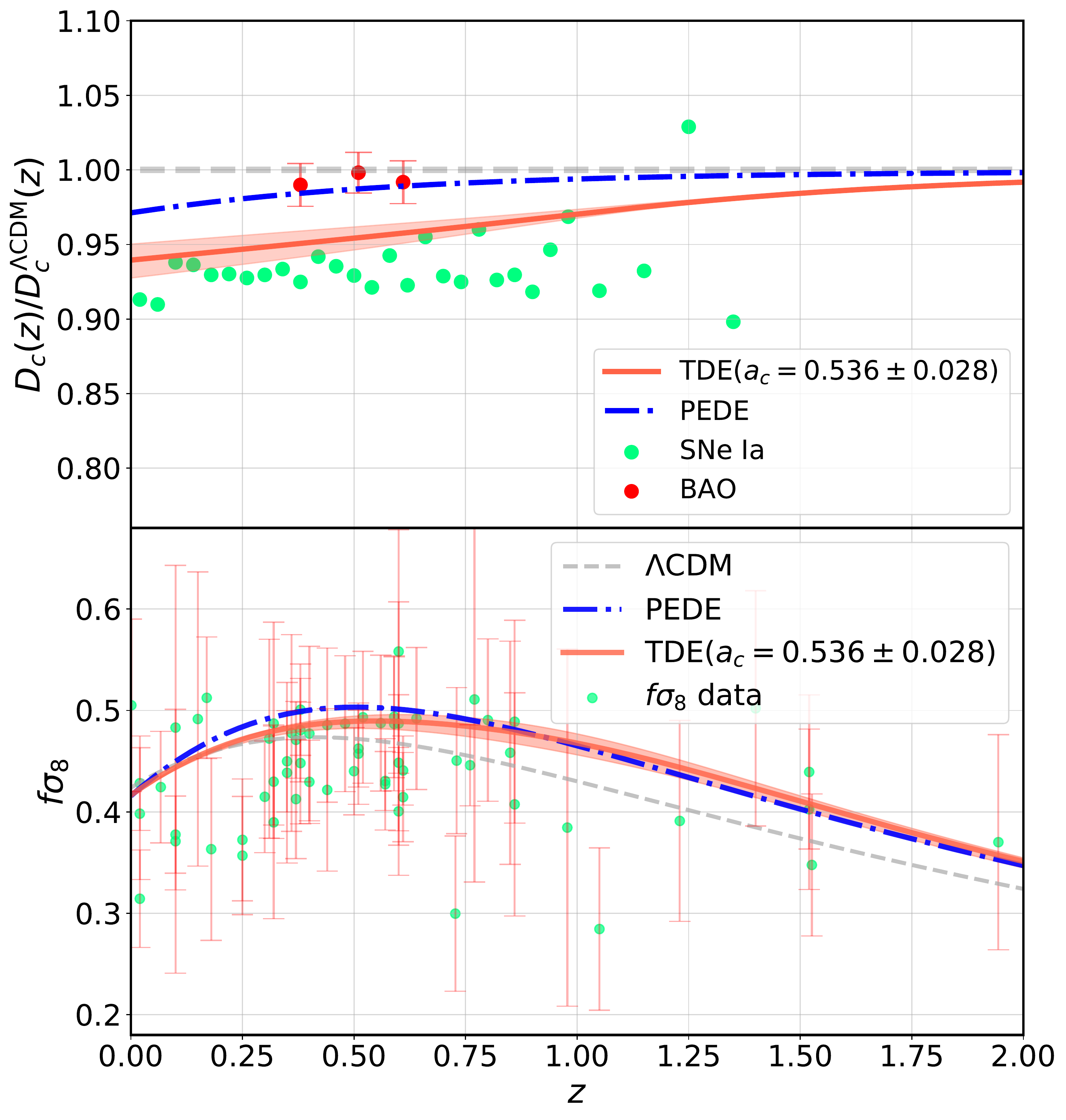}
	\caption{The relative comoving distances ($D_c(z)/D^{\Lambda\rm{CDM}}_c(z)$)
		are shown in the top panel, and $f(z)\sigma_{8}(z)$ are shown in the bottom panel, respectively.
		The cosmological parameters are set to the best-fit values of joint \emph{Planck}+$M_{\rm{B}}$+SNe datasets.
		The SNe constrain only the ratio of the distances $D_L /D_{H_0}$, so the absolute scale of the SNe data points in the top panel is calibrated by the $M^{\rm{R20}}_{\rm{B}}$ prior (\citet{Efstathiou:2021ocp}), i.e., $M_{\rm{B}} = -19.244\pm 0.037$.  The BAO measurements constrain $D_c\times(r_{\rm{d,fld}}/r_{\rm{d}})$, and thus the BAO data point are corrected with the factor $r_{\rm{d}}/r_{\rm{d,fld}}$($r_{\rm{d,fld}}=147.78$). Noting that, the inferred values of $r_d$ are the same for PEDE  (\citet{Li:2019yem}) and TDE model (when $\omega_m$ and $\omega_r$ are fixed).  		
		The 63 observational $f\sigma_{8,obs}(z)$ RSD data points are collected by (\citet{Kazantzidis:2018rnb}), which have been corrected by the AP factor (\citet{Li:2019nux}) calculated based on TDE model.
	}\label{fig:lss_fs8}
\end{figure}

\section{MODEL}\label{sec:models}

According to the recent surveys, the universe on large scale is almost flat and can be well described by Friedmann-Lemaitre-Robertson-Walker (FLRW) metric.  The Hubble parameter within this framework can be 
expressed as:
\begin{eqnarray}
H(z) = H_0\left[\Omega_{m,0}(1+z)^3+\Omega_{r,0}(1+z)^4+\Omega_{\rm{DE}}\right]^{\frac{1}{2}},
\end{eqnarray}
where $\Omega_{m,0},\Omega_{r,0}$ are the present time matter and radiation density, respectively. The dark 
energy density $\Omega_{\rm{DE}}$ can be described as:
\begin{eqnarray}
\Omega_{\rm{DE}}(z) = \Omega_{\rm{DE},0} \times \exp\left[3\int^{z}_{0}\frac{1+w(z')}{1+z'}dz\right],
\label{eq:de_denstiy}
\end{eqnarray}
where $\Omega_{\rm{DE}}$ is ratio of DE density to the critical density ($\rho_{\rm{DE}}/\rho_{crit,0}$). 
The exponential part of Eq.\ref{eq:de_denstiy} indicates that, 
a smaller value of $w$ would translate into a faster dilution rate of DE.
This effect suppresses the dilation of the late universe and is well compensated by an upward shift in $H_0$, thus, the $H_0$ tension is relieved.

\subsection{Transitional Dark Energy}
We formulate a novel version of transitional dark energy (TDE) model (see (\citet{Keeley:2019esp,Bassett:2002qu,Shafieloo:2009ti}) for discussions of similar models), with the EoS of DE 
written as:
\begin{eqnarray}
w(z) = w_{0} -\frac{1}{2}\left[\tanh\left(3\left(\frac{1}{a}-\frac{1}{a_c}\right)\right)+1\right].
\end{eqnarray}
To recover $\Lambda$CDM model when $a \gtrsim a_c,$ we fix $w_0 = -1$ (denote as TDE(1p)).
In such case we have the EoS of DE evolving from $w = -2$ at high redshift
to $w = -1$ at low redshift, which recovers the case in Covariant Galileon cosmology under the right initial conditions (\citet{DeFelice:2010pv}). As is illustrated in Fig. \ref{fig:w_fld}
the EoS of DE remains constant at late-time then endures a rapid transition at the critical scale $a_c$.
With the increase of $a_c$, the inferred value of $H_0$ increases, while the shifts on $S_8$ is the reverse. However, our analysis suggests that the 
transition of DE could not relieve the $f\sigma_8$ tension (see Fig. \ref{fig:lss_fs8}). 
Owing to the fact that, the growth function $f(z)$ is boosted at $a\lesssim a_c$, thus the solution of $H_0$ and 
$S_8$ tension (raising $a_c$) is at the cost of exacerbating $f\sigma_8$ tension.

Due to the resemblance to $\Lambda$CDM model 
at late universe, TDE(1p) could largely avoid the mismatch with SNe datasets at the same time fit with 
the acoustic scale determined by CMB. \minew{However, this scenario does not 
relieve the mismatch between the local distance ladder and BAO surveys, both of which 
can be directly related to the supernova absolute magnitude $M_B$.
With the calibration of the sound horizon $r_s(z_d)$ (determined by CMB), the BAO experiments (\cite{Alam:2016hwk}) anchor the distance scales ($D_{A}(z\sim0.3)$) at the late-universe.
Using the parametric-free inverse distance ladder (\cite{Camarena:2019rmj})
these scales can be propagated to SnIa absolute magnitude:
\begin{eqnarray}
M^{\rm{P18}}_{\rm{B}} = -19.401 \pm 0.027 \: \rm{mag},
\end{eqnarray}
which disagrees with the $M_B$ prior obtained from SH0ES (\citet{Riess:2020fzl}) by demarginalization method at $3.4\sigma$ (\citet{Camarena:2019moy}). Note that $r_s(z_d)$ is preserved within the scenario of late-DE.
In this light, to address the above discussed $M_B$ tension, one may consider a rapid phantom transition of the EoS of DE
at $z<0.1$ accompanied by a similar transition in the value of $M_{\rm{B}}$ (\citet{Alestas:2020zol,Marra:2021fvf}).
} 

\begin{figure}
	\centering
	\includegraphics[scale=0.3]{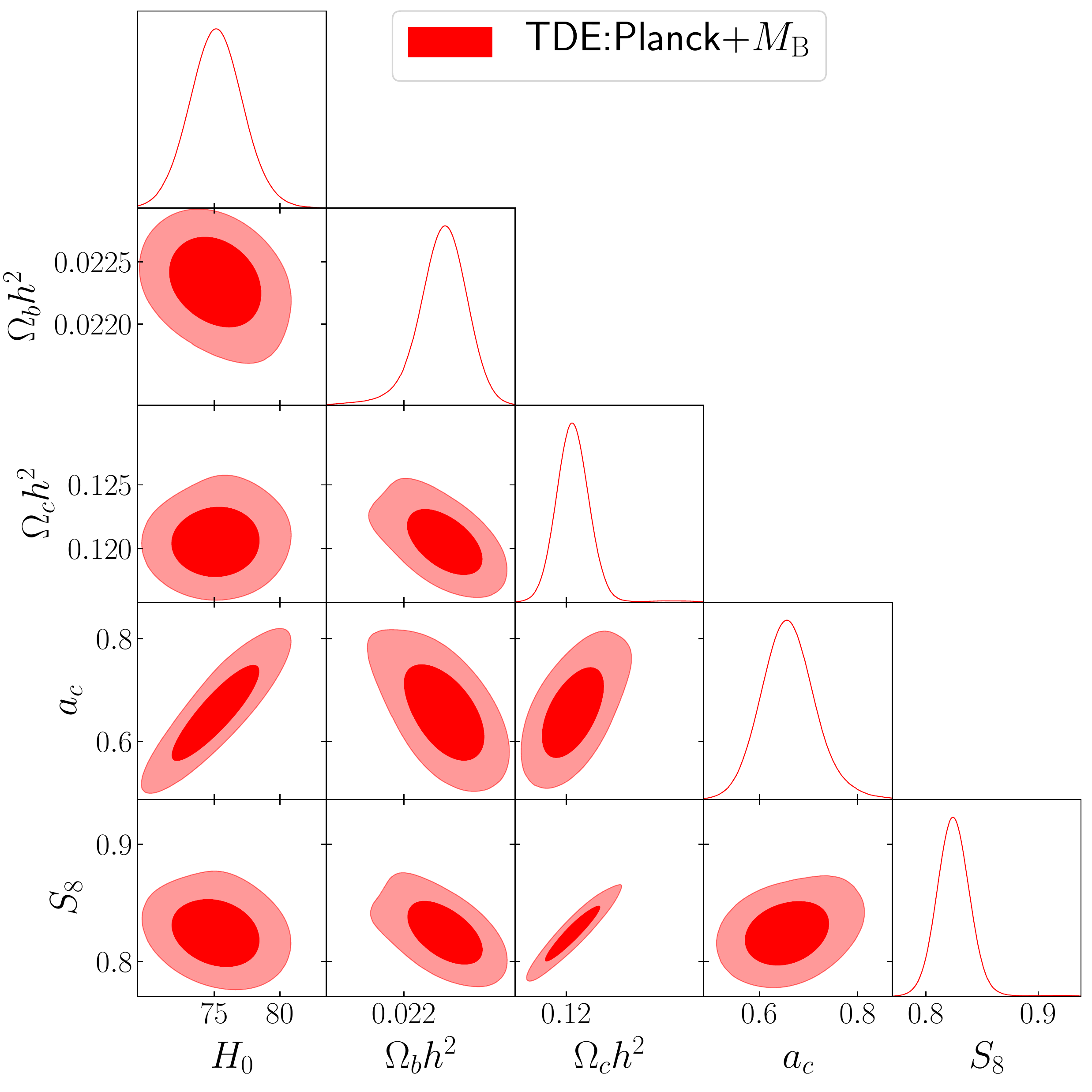}
	\caption{Cosmological parameter constraints from the combination of \emph{Planck} 2018 distance prior and the $M^{\rm{R20}}_{\rm{B}}$ prior; the contours show $1\sigma$ and $2\sigma$ posteriors for the TDE model. }
	\label{fig:plc_H0_contours}
\end{figure}

\begin{table}
	\renewcommand{\arraystretch}{1.2}
	\begin{center}
		\begin{tabular}{|cc|}
			\hline
			\hline
			Parameter                    & Prior\\
			\hline 
			\hline
			$H_0$                        & $[50,90]$ \\
			$\Omega_{b} h^2$             & $[0.01,0.1]$\\
			$\Omega_{c} h^2$             & $[0.05,0.3]$\\
			$a_c$(TDE)                   & $[0.01,1.0]$\\
			\hline
		\end{tabular}
	\end{center}
	\caption{Flat priors assumed on the cosmological parameters associated with the TDE model and $\Lambda$CDM model}
	\label{tab:priors}
\end{table}

\section{STATISTICAL METHODOLOGY AND DATASETS}\label{sec:data}
We implement TDE scenario as modifications to
the publicly available Einstein-Boltzmann code \href{http://class-code.net/}{\texttt{CLASS}} (\citet{Lesgourgues:2011re,Blas:2011rf}) package.
The non-linear matter power spectrum required by redshift-space distortion (RSD) likelihoods are computed using the ``HMcode'' (\citet{Mead:2015yca,Mead:2016ybv,Mead:2020vgs}) implemented in \texttt{CLASS}.
The MCMC analyses are preformed 
using the publicly available code \href{https://cobaya.readthedocs.io}{\texttt{Cobaya}} (\citet{Torrado:2020dgo}) package
with a Gelman-Rubin (\citet{10.1214/ss/1177011136}) convergence
criterion $R-1 < 0.05$. The plots have been obtained
using the \href{https://getdist.readthedocs.io}{\texttt{GetDist}} (\citet{Lewis:2019xzd}) package.

The following datasets are considered in the MCMC analyses:
\begin{table*}
	\renewcommand{\arraystretch}{1.4}
	\begin{center}
		\begin{tabular}{|llllllllll|}			
			\hline
			\hline			
			Model\hspace{2pt} &Dataset & $H_0$ & $S_8$ & $a_c$ &  $\sigma_{\rm SH0ES}$& $\chi^2_{\rm{min}}$&$\Delta\chi^2$ &$\ln\varepsilon$&$\Delta\ln\varepsilon$ \\
			\hline
			\hline
			
			TDE & \emph{Planck}  &  $77.0\pm 8.0   $ & $0.815^{+0.032}_{-0.027}  $ & $> 0.562$ &$0.2$ & $2.9$ &0.0 &-8.73 &0.69\\
			& \emph{Planck} + $M_{\rm{B}}$  &  $75.2\pm 1.7$ & $0.825\pm0.014$ & $0.658\pm0.049$ & 0.1& $4.6$ &-23.5 &-11.54&-9.10 \\
			&PMS (\emph{Planck} +  $M_{\rm{B}}$ +SNe)  &  $71.59\pm 0.81 $ & $0.825\pm 0.011 $ & $0.548\pm 0.029$ & 2.0& $1045.0 $ &-17.5&-533.03&-5.10\\
			&PMS + $S_8$ & $71.77\pm0.90   $ & $0.812\pm0.012$ & $0.531^{+0.045}_{-0.040}$ & 1.9& $1051.3 $ &-14.7&-535.07&-4.06\\
			&PMS + $S_8$ +BAO & $69.55^{+0.73}_{-0.94}$ & $0.816\pm 0.015$ & $0.420\pm0.082$ & 3.0& $1062.3 $ &-5.6&-540.25&-0.89\\
			&PMS + $S_8$ +BAO +FS & $69.16\pm 0.76 $ & $0.814\pm 0.012 $ & $0.375^{+0.12}_{-0.043}  $ & 3.2& $1068.2 $&-3.5&-543.41&0.15\\
			\hline
			\hline
			& & $H_0$ & $S_8$ & $-$ & $\sigma_{\rm SH0ES}$&  $\chi^2_{\rm{min}}$&$\Delta\chi^2$ &$\ln\varepsilon$&$\Delta\ln\varepsilon$\\
			\hline
			LCDM& \emph{Planck}   &  $66.77\pm 0.54  $ & $0.836\pm 0.013$ & $-$ & 4.9& 2.9    & $- $&-8.04&$-$\\
			& \emph{Planck} +  $M_{\rm{B}}$  &  $67.41^{+0.46}_{-0.65}  $ & $0.820^{+0.015}_{-0.011}$ & $-$ & 4.5& 28.1  & $- $&-20.64&$-$\\
			&PMS   &  $67.74\pm 0.50       $ & $0.813\pm 0.012  $ &$ -$ & 4.3& 1062.5 & $- $&-538.13&$-$\\
			&PMS + $S_8$   &  $68.34^{+0.76}_{-0.33}  $ & $0.798^{+0.014}_{-0.020} $ & $- $ & 4.0& 1066.0 & $- $ & -539.13&$-$\\
			&PMS + $S_8$ +BAO & $68.26^{+0.59}_{-0.38}   $ & $0.800^{+0.013}_{-0.015}  $ & $- $ & 4.0&  1067.9 & $- $&-541.14&$-$\\
			&PMS + $S_8$ +BAO +FS & $68.35\pm 0.31 $ & $0.800\pm 0.010 $ & $- $ & 4.1 &1071.7& $- $&-543.26&$-$\\
			\hline
			
		\end{tabular}
	\end{center}
	\caption{The mean and $1$-$\sigma$ constraints on $H_0$ and $S_8$ in the TDE model (top panel) and $\Lambda$CDM model (bottom panel), as well as the the constraints on $a_c$ for TDE model. We have also shown in the third to the last column the effective number of $\sigma$'s that the Hubble measurement is away from SH0ES. The $\chi^2$ statistics are shown in the last two columns, where $\Delta \chi^2$ is the
		inferred $\chi^2_{\rm{min}}$ compared with $\Lambda$CDM model. In the last column, $\Delta\ln\varepsilon$ denotes difference of Bayesian evidence between $\Lambda$CDM model and the TDE model, i.e., $\Delta\ln\varepsilon = \ln\varepsilon_{\Lambda\rm{CDM}}-  \ln\varepsilon_{\rm{TDE}}$. }
	\label{tab:accumulate_constraints}
\end{table*}

\subsection{CMB}
We consider the CMB distance prior from final \emph{Planck} 2018 release. Noted that, in late-DE scenarios, the shape of the power spectrum (with the same matter and radiation density) is identical to that of $\Lambda$CDM cosmology at $\ell \gtrsim 10$ (for small $\ell s$ the covariance is largely dominated by cosmic variance). In this case, the fit of the distance prior is equivalent to that of the angular power spectrum, while this is not true for EDE scenarios. 
\minew{The analysis with the \emph{Planck} 2018 low-$\ell$ TT+EE and \emph{Planck} 2018 high-$\ell$ TT+TE+EE temperature and polarization power spectrum (\cite{Planck:2019nip}) are shown in Appendix \ref{sec:planck}} 
\subsection{Hubble Constant}
The most recent SH0ES measurement indicates that $H_0 = 73.2 \pm
1.3$ (\citet{Riess:2020fzl}), which shows a tension at $4.4\sigma$ with the \emph{Planck} (\citet{Aghanim:2018eyx}) value of
$H_0$ assuming a minimal $\Lambda$CDM model.
\minew{Note that the determination of $H_0$ by the SH0ES collaboration have fixed the deceleration parameter to the standard $\Lambda$CDM model value of $q_0 = -0.55$. To avoid inconsistency and double counting of the low-z SNe samples (\citet{Efstathiou:2021ocp}), we adopt a prior on the SNe Ia absolute magnitude
\begin{eqnarray}
M^{\rm{R20}}_{\rm{B}} = -19.244\pm 0.037 \: \rm{mag}
\end{eqnarray} 
which is derive using the demarginalization methodology of \citet{Camarena:2019moy}.
With an uninformative flat prior for $q_0$, the above $M_{\rm{B}}$ prior gives $H_0 = 75.35\pm 1.68$.
We denote as $\sigma_{\rm SH0ES}$ the effective number of $\sigma's$ that the Hubble measurement is away from this value.
See \citet{Camarena:2021jlr} for a detailed discussion of the local $M_{\rm{B}}$ likelihoods.
}

\subsection{Supernovae Type Ia}
The 1048 Supernoave Type Ia data points
distributed in redshift interval $z \in [0.01, 2.3]$, known as the Pantheon sample (\citet{Scolnic:2017caz}), which provide accurate relative luminosity distances. We utilize the likelihood as implemented in Cobaya.

\subsection{LSS}
The LSS dataset which probes the low-redshift universe are considered, which include:
\begin{itemize}
	\item BAO: measurements of the BAO signal with their full covariance
	matrix from BOSS DR12 (\citet{Alam:2016hwk}). Radial and transverse BAO measurements from Ly$\alpha$ Forests SDSS DR12 (\citet{Bautista:2017zgn}). Angle averaged BAO measurement from 6DF Galaxy (\citet{Beutler:2011hx}), 
	from quasar sample from BOSS DR14 (\citet{Gil-Marin:2018cgo}).
	\item RSD: 
	SDSS BOSS DR12 (\citet{Alam:2016hwk}) measurements of $f\sigma_8(z)$, at $z =0.38$, 0.51 and 0.61. We include the
	full covariance of the joint BOSS DR12 BAO and
	RSD data (denote as FS). Noting that the measured $f\sigma_8(z)$ at different redshfts is not fully independent. Also, the $f\sigma_8(z)$ data is correlated with the BAO signal, thus, it's hard obtain the full 
	covariance of the datasets compiled from different surveys. In view of this, the compilation of 63 $f\sigma_{8,\rm{obs}}(z)$ RSD data points (\citet{Kazantzidis:2018rnb}) are not included in the MCMC analysis, but only displayed as an illustration of the $f\sigma_8$ tension.
	\item DES:
	shear-shear, galaxy-galaxy, and galaxy-shear two-point correlation functions (“3x2pt”), 
	measured from 26 million source galaxies in
	four redshift bins and 650,000 luminous red lens galaxies
	in five redshift bins, for the shear and galaxy correlation
	functions, respectively. We utilize a Gaussian prior on $S_8$ derived by the DES-Y1 likelihood as is suggested in (\citet{Hill:2020osr}).
\end{itemize}

It's worth noting that TDE are extended from $\Lambda$CDM model, and thus share with $\Lambda$CDM model the six parameters. We fix three of the six parameter ($n_s$,$A_s$,$\tau_{\rm{reio}}$) associated with the overall shape of the CMB spectrum to the best-fit value given by \emph{Planck} 2018 (\citet{Aghanim:2018eyx}).
Next, we consider as baseline a 4-dimensional parameter space described by the following parameters: the Hubble constant $H_0$, the baryon
energy density $\Omega_{b}h^2$, cold dark matter energy density $\Omega_{c}h^2$, and the transitional scale $a_c$ for TDE. We assume
flat uniform priors on all these parameters, as are shown in Table \ref{tab:priors}.

\begin{figure*}
	\centering
	\includegraphics[scale=0.4]{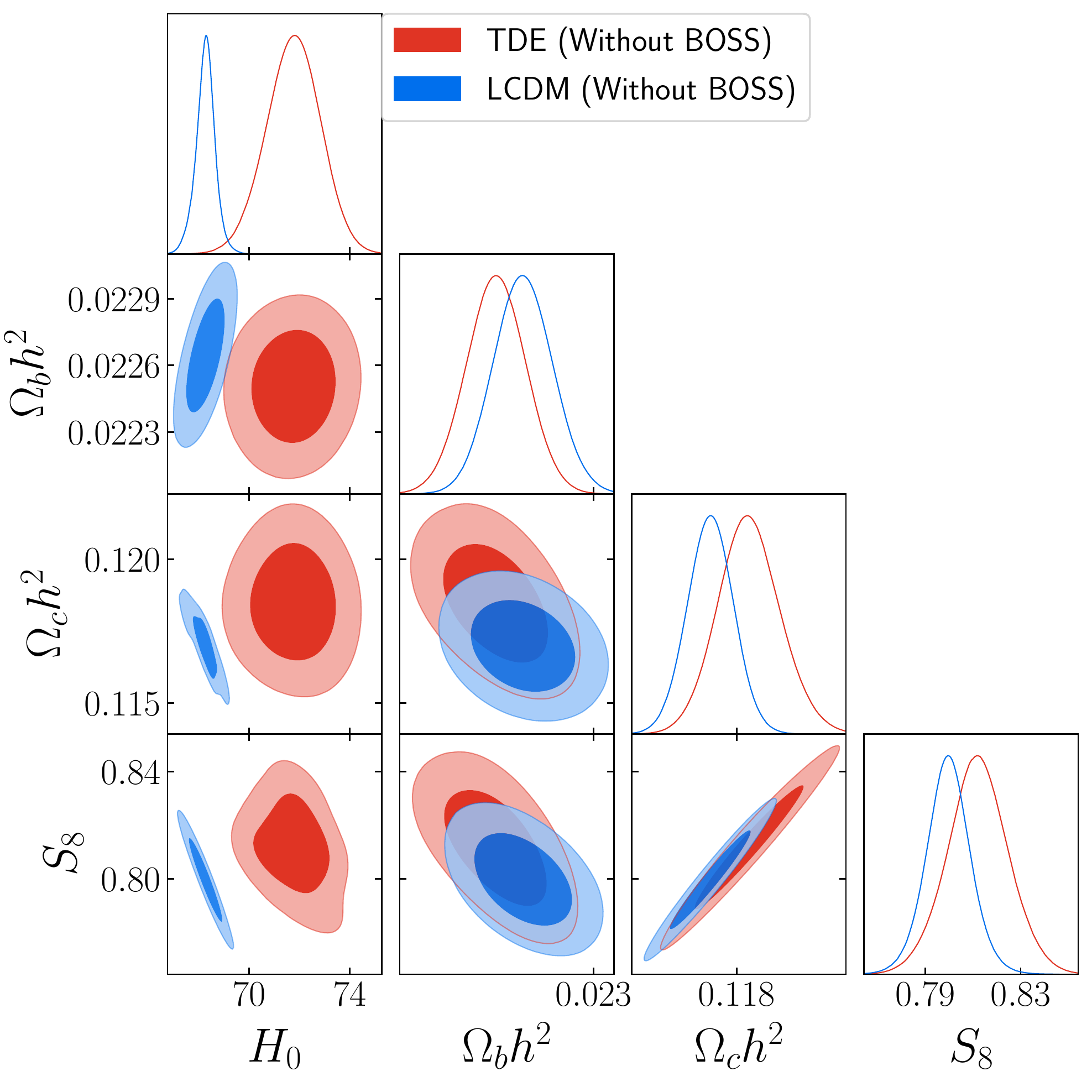}
	\includegraphics[scale=0.4]{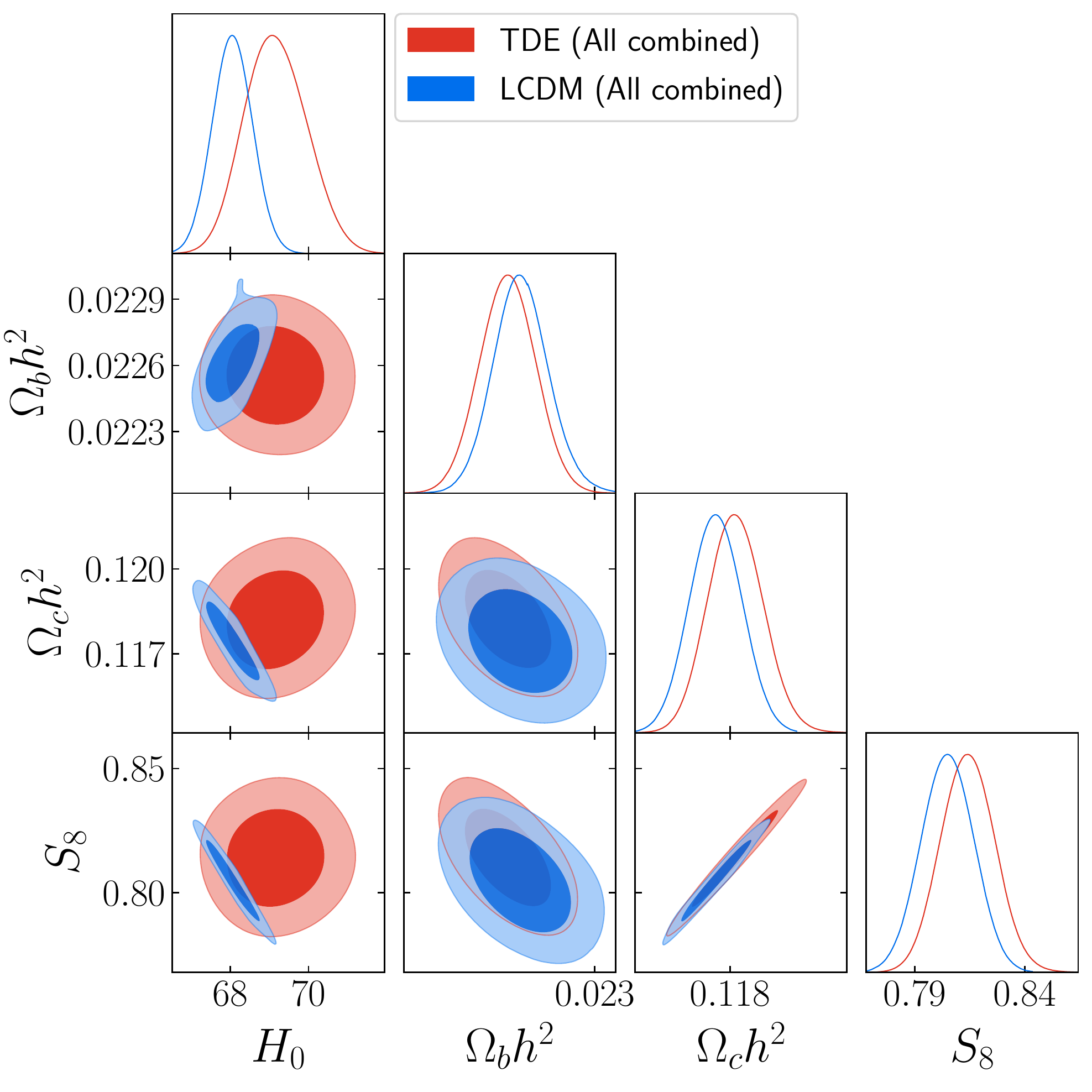}
	\caption{The right panels show the cosmological parameters constraints from the combination of \emph{Planck} 2018 distance prior; BAO data from 6dF, SDSS DR14, and BOSS DR12; Pantheon SNIa data; the $M^{\rm{R20}}_{\rm{B}}$ prior; BOSS DR12 RSD data; and the $S_8$ prior derived from DES-Y1 3x2pt data. The red (blue) contours show $1\sigma$ and $2\sigma$ posteriors for the TDE($\Lambda$CDM) model.  
		In the left panel, we show the show $1\sigma$ and $2\sigma$ posteriors for the TDE($\Lambda$CDM) model by constraints from the above combination of without the full covariance of the joint BOSS DR12 BAO and RSD data (denote as without BOSS); The posteriors for all datasets combinations are shown in the right panel as a contrast (denote as all combined).
	}
	\label{fig:selected_contours}
\end{figure*}

\begin{figure}
	\centering
	\includegraphics[scale=0.35]{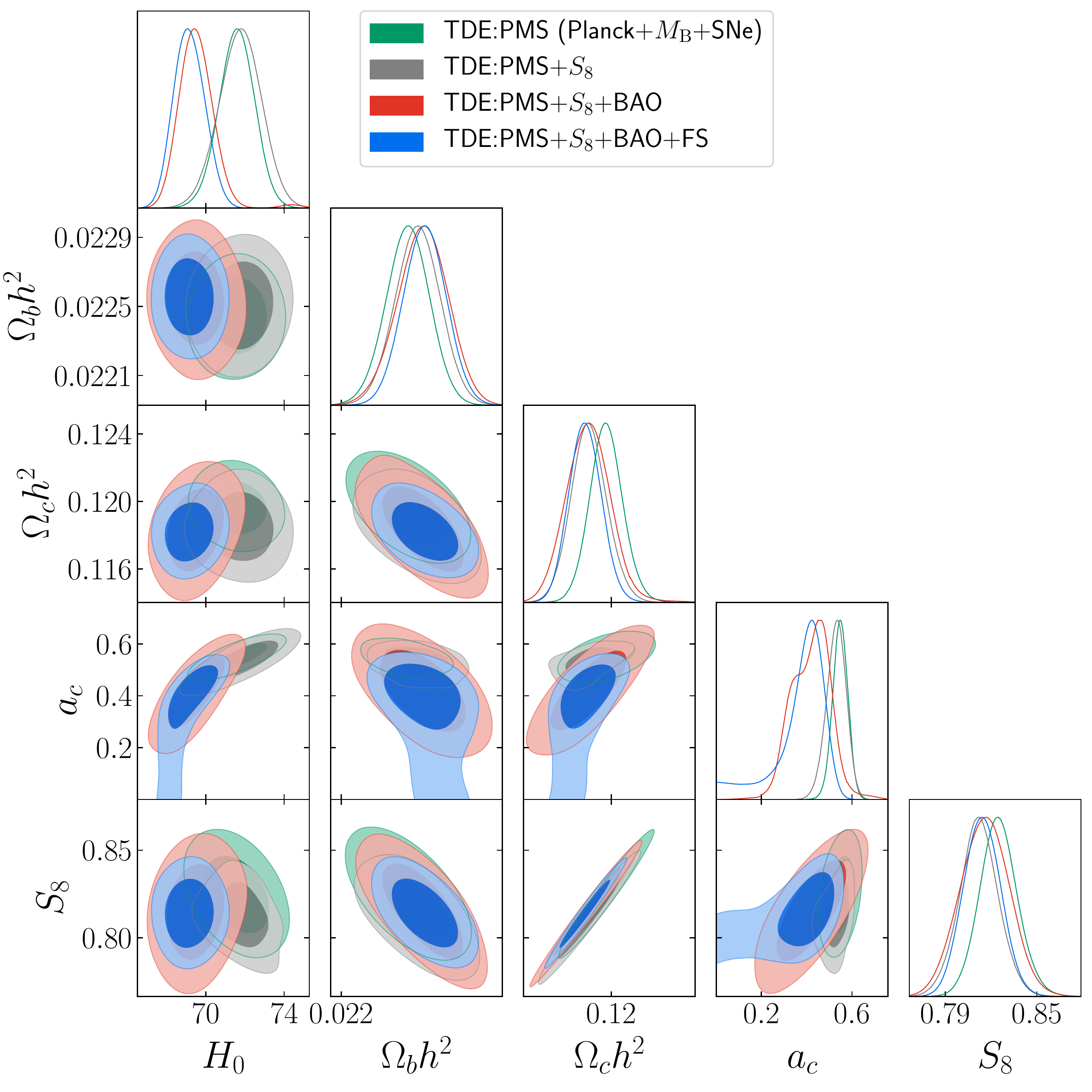}
	\caption{Constraints on TDE scenario from various dataset combinations. The contours show $1\sigma$ and $2\sigma$ posteriors. 
		The green contours show results for the combination \emph{Planck} distance prior, SNIa and $M_{\rm{B}}$ prior; the dark contours additionally include a prior on $S_8$; and the red contours further include BAO data; the blue contours add the RSD data (constrains on $f\sigma_{8}$).	
		Noting that, the inclusion of a prior on $S_8$ slightly alleviate the tension with SH0ES and shift $a_c$ toward 1.}
	\label{fig:TDE_MEDE_contours}
\end{figure}

\begin{table}
	\renewcommand{\arraystretch}{1.5}
	\centering
	\begin{tabular}{|lcc|}
		\hline\hline Dataset &\multicolumn{2}{c|}{\emph{Planck}+ $M_{\rm{B}}$ }    \\ \hline
		Model &~~~ TDE  ~~~&$\Lambda$CDM~~\\ \hline \hline
		{\boldmath$\rm{H}_0      $} & $75.2\pm 1.7            $ & $67.41^{+0.46}_{-0.65}             $\\
		
		{\boldmath$\Omega_bh^2    $} & $0.02233^{+0.00016}_{-0.00015}   \:      $ & $0.02248^{+0.00012}_{-0.00017}      \:      $ \\
		
		{\boldmath$\Omega_ch^2    $} & $0.1206^{+0.0011}_{-0.0013}          $ & $0.1186^{+0.0012}_{-0.0011}           $\\
		
		{\boldmath$ a_{c}          $} & $0.658\pm 0.049      $&  $-$\\

		$S_8                        $ & $0.825^{+0.013}_{-0.015}            $& $0.820^{+0.015}_{-0.011}         $\\
		
		$\chi^2_\mathrm{M_B}     $ & $0.8\pm 1.2                 $ & $24\pm 4          $\\
		
		$\chi^2_\mathrm{\emph{Planck}}	   $ & $3.9\pm 6.4                 $ &  $4.2\pm 3.4             $\\

		$\chi^2_{tot}			   $ & $4.6\pm 6.6    $  & $28.1\pm 1.9          $\\
		\hline
		$\Delta\chi^2			   $ & $-23.5 		   $  &  $-$ \\
		$\Delta\rm{AIC}			   $ & $-21.5 		   $  &  $-$ \\
		$\Delta\ln\varepsilon		   $ & $-9.10 		   $ &  $-$ \\
		\hline
	\end{tabular}
	\caption{The mean $\pm1\sigma$ constraints on the cosmological parameters in $\Lambda$CDM and in the TDE scenario, as inferred from the combination of \emph{Planck} 2018 distance prior and the $M^{\rm{R20}}_{\rm{B}}$ prior. Noting that, $\Delta\ln\varepsilon = \ln\varepsilon_{\Lambda\rm{CDM}}- \ln\varepsilon_{\rm{TDE}}$.}
	\label{tab:plc_H0}
\end{table}

\begin{table*}
	\renewcommand{\arraystretch}{1.4}
	\centering
	\begin{tabular}{|lcc|cc|}
		\hline\hline Dataset &\multicolumn{2}{c|}{Without BOSS }  &\multicolumn{2}{c|}{Without SNe }  \\ \hline
		Model &~~~ TDE  ~~~&$\Lambda$CDM~~ &~~~ TDE  ~~~&$\Lambda$CDM~~\\ \hline \hline
		{\boldmath$\rm{H}_0       $} & $71.78\pm 0.90             $ & $68.34^{+0.76}_{-0.33}             $
		& $69.55^{+0.89}_{-1.1}          $ & $68.17\pm 0.39             $\\
		
		{\boldmath$\Omega_bh^2    $} &  $0.02250\pm 0.00013    \:    $& $0.02268^{+0.00018}_{-0.00014} \:           $
		& $0.02251^{+0.00016}_{-0.00013}   \:     $ & $0.02266^{+0.00013}_{-0.00014} \:           $ \\ 
		
		{\boldmath$\Omega_ch^2    $} & $0.1185\pm 0.0011         $ & $0.1169^{+0.0012}_{-0.0016}         $
		&$0.1187^{+0.0011}_{-0.0014}          $ & $0.11725\pm 0.00077              $\\
		
		{\boldmath$ a_{c}          $} & $0.531^{+0.045}_{-0.040}    $& $-             $
		& $0.419^{+0.11}_{-0.057}              $& $-             $\\

		$S_8                        $ & $0.812\pm 0.012            $& $0.7996\pm 0.0087          $
		& $0.819^{+0.011}_{-0.014}            $& $0.8023\pm 0.0091          $\\
		
		$\chi^2_\mathrm{M_B}     $ & $4.8\pm 2.4                 $ &   $18.0\pm 4.2                $
		& $11.5\pm 4.6                  $  & $18.8\pm 2.4            $\\
		
		$\chi^2_\mathrm{\emph{Planck}}	   $ & $4.9\pm 3.3                 $ & $11.5\pm 5.1           $
		& $4.9\pm 3.8               $ &  $9.1\pm 4.1         $\\
		
		
		$...$ & & & &\\
		
		$\chi^2_{tot}			   $ & $1051.3\pm 2.8  $ & $1066.0\pm 3.9            $
		& $34\pm 6  $  & $37.4\pm 2.5             $\\
		\hline
		$\Delta\chi^2			   $ & $-14.7 		   $  &  $-$ 
		& $-3.4 		   $ &   $-$ \\
		$\Delta\rm{AIC}			   $ & $-12.5 		   $ &  $-$ 
		& $-1.4 		   $ &  $-$ \\
		$\Delta\ln\varepsilon			 $ & $-4.06 		   $ &   $-$ 
		& $0.23 		   $  &  $-$ \\
		\hline
	\end{tabular}
	\caption{The the mean $\pm1\sigma$ constraints on the cosmological parameters in TDE and $\Lambda$CDM models, as inferred from the combination of \emph{Planck} 2018 distance prior; BAO data from 6dF; Pantheon SNIa data; the $M^{\rm{R20}}_{\rm{B}}$ prior; the $S_8$ prior derived from DES-Y1 3x2pt data (left panel), and the
		\emph{Planck} 2018 distance prior; BAO data from 6dF, SDSS DR14;  $M_{\rm{B}}$ prior; and the $S_8$ prior (right panel).  $\Delta \rm{AIC}$ and $\Delta \chi^2$ are the differences be AIC criteria and $\chi^2_{\rm{min}}$ inferred in TDE model with that of $\Lambda$CDM model, respectively. We show in the last row the pair difference of Bayesian evidence: $\Delta\ln\varepsilon = \ln\varepsilon_{\Lambda\rm{CDM}}-  \ln\varepsilon_{\rm{TDE}}$. }
	\label{tab:BOSS_SNe}
\end{table*}

\begin{table*}
	\renewcommand{\arraystretch}{1.4}
	\centering
	\begin{tabular}{|lcc|cc|}
		\hline\hline Dataset &\multicolumn{2}{c|}{Without  $M_{\rm{B}}$ } &  \multicolumn{2}{c|}{All combined }  \\ \hline
		Model &~~~ TDE  ~~~&$\Lambda$CDM~~&~~~ TDE  ~~~&$\Lambda$CDM~~\\ \hline \hline
		{\boldmath$\rm{H}_0       $} & $67.82^{+0.48}_{-0.57}            $ &  $67.67\pm 0.43           $
		& $69.16\pm 0.76               $ &  $68.06\pm 0.41           $\\
		
		{\boldmath$\Omega_bh^2    $} & $0.02257\pm 0.00014  \:      $ & $0.02260\pm 0.00013  \:          $ 
		&$\:0.02255\pm 0.00013    \:    $ & $0.02264^{+0.00013}_{-0.00011}      \:      $ \\
		
		{\boldmath$\Omega_ch^2    $} & $0.1180\pm 0.0009          $ & $0.1181\pm 0.0009            $
		& $0.1182\pm 0.0010        $& $0.1174\pm 0.00086            \:$\\
		
		{\boldmath$ a_{c}          $} & $< 0.252         $& $-             $
		& $0.375^{+0.12}_{-0.043}      $ & $-            $\\

		$S_8                        $ & $0.818\pm 0.011           $& $0.814\pm 0.010          $
		& $0.814\pm 0.012            $& $0.805\pm 0.010          $\\
		
		$\chi^2_\mathrm{\emph{Planck}}	   $ & $4.9\pm 3.0               $  & $5.5\pm 3.1           $
		& $5.6\pm 3.5                  $  & $9.9\pm 3.2     $\\
		$\chi^2_\mathrm{SN}	       $ & $1034.99\pm 0.24                $ & $1034.99\pm 0.23$
		& $1068.4\pm 2.9                 $& $1034.80\pm 0.06    $ \\
		
		$...$ & &  && \\
		
		$\chi^2_{tot}			   $ & $1051.8\pm 2.7   $ & $1051.9\pm 2.2            $
		& $1067.2\pm 2.9  $ & $1071.7\pm 2.2          $\\
		\hline
		$\Delta\chi^2			   $ & $-0.1 		   $  &  $-$ &$-4.5 		   $  &  $-$ \\
		$\Delta\rm{AIC}			   $ & $ 1.9 		   $  &  $-$  & $-2.5 		   $  &  $-$ \\
		$\Delta\ln\varepsilon		$ & $ 2.74 		   $  &  $-$  & $0.15 		   $  &  $-$ \\
		\hline
	\end{tabular}
	\caption{The the mean $\pm1\sigma$ constraints on the cosmological parameters in TDE and $\Lambda$CDM models, as inferred from the combination of \emph{Planck} 2018 distance prior; BAO data from 6dF, SDSS DR14, and BOSS DR12; Pantheon SNIa data; BOSS DR12 RSD data; and the $S_8$ prior derived from DES-Y1 3x2pt data (left panel), and \emph{Planck} 2018 distance prior; BAO data from 6dF, SDSS DR14;  $M_{\rm{B}}$; SNe dataset and the $S_8$ prior (right panel). $\Delta \rm{AIC}$ and $\Delta \chi^2$ are the differences between
		 the value of AIC criteria and $\chi^2_{\rm{min}}$ inferred in TDE model with that of $\Lambda$CDM model, respectively. The pair difference of Bayesian evidence given in the last row is: $\Delta\ln\varepsilon = \ln\varepsilon_{\Lambda\rm{CDM}}-  \ln\varepsilon_{\rm{TDE}}$. }
	\label{tab:SH0ES_combined}
\end{table*}


\section{RESULTS AND DISCUSSION}\label{sec:results}
For the cosmological analysis, we fit the TDE model to different combinations of the above discussed datasets.  
As a comparison, we also include the analysis of the $\Lambda$CDM model. 
In order to compare among these models, we compute the Bayesian evidence $\varepsilon$, which is a crucial quantity for model selection
and has been widely used in cosmology (\citet{Mehrabi:2018oke,Rezaei:2020mrj,Lonappan:2017lzt,Saini:2003wq}). Further, given two models $M_i$ and $M_j$ one can use the Jeffreys’ scale (\citet{jeff.book}) to measure the significant difference between two models, i.e., $\Delta\ln\varepsilon = \ln\varepsilon_{M_i}-\ln\varepsilon_{M_j}$. The model pair difference provides the following 
situations:
\begin{itemize}
	\item $0<\Delta\ln\varepsilon<1.1$ suggests weak evidence against $M_j$ model when compared with $M_i$.
	\item for $1.1 <\Delta\ln\varepsilon< 3$  there is a definite evidence against	model $M_2$.
	\item for $3 <\Delta\ln\varepsilon$  there is a strong evidence against model $M_2$.
\end{itemize}
Correspondingly, the evidence is against $M_i$ model when $\Delta\ln\varepsilon<0$. 
In addition to the evidence, we also include the computation of the Akaike Information (AIC) (\citet{Akaike:1974}), given by:
\begin{eqnarray}
\rm{AIC} = \chi^2_{\rm{min}}+2k
\end{eqnarray} 
where k is the number of fitting parameters.
For brevity, we denote the combination of likelihoods without LSS data as PMS (\emph{Planck}+ $M_{\rm{B}}$+SNe).
In Table \ref{tab:accumulate_constraints} we report the constraints at 68\% CL on the $H_0$ and some key derived quantities for several datasets combinations considered in this work. Detailed constrains on cosmological parameters and $\chi^2$ statistics can be found in Tables \ref{tab:plc_H0}$\sim$\ref{tab:SH0ES_combined}. The triangular plot with the 1D posterior distributions and the 2D contour plots for these parameters are shown in Fig. \ref{fig:plc_H0_contours}$\sim$\ref{fig:TDE_MEDE_contours}.

\subsection{Constraints from CMB and SH0ES}
As is discussed in section \ref{sec:ph}, there is a strong degeneracy between $H_0$ and the late-DE parameter $a_c$ ($w$) which quantify the deviation of these scenarios from $\Lambda$CDM model (see Fig. \ref{fig:shooting_results}).
Consequently, the \emph{Planck} data alone is insufficient to give a tight constrain to these parameters, so that a prior on $H_0$ should be included.
To test the above argument, we first consider the fit to \emph{Planck} 2018 distance prior alone. The results tabulated in Table
\ref{tab:accumulate_constraints} indicates a lower bound of $a_c>0.562$ at $68\%$ CL, while $H_0$ is loosely constrained. 
Next, we consider the fit to a combination of \emph{Planck} distance prior and the $M^{\rm{R20}}_{\rm{B}}$ prior derived from SH0ES. In Table \ref{tab:plc_H0}, we find the Hubble constant to be $H_0 = 75.2\pm1.7$.
The $H_0$ tension in TDE scenario is largely removed when the $M^{\rm{R20}}_{\rm{B}}$ prior is added. Noting that, the resolve of the tension between local distance ladder and CMB can be achieved by a wide class of ``phantom-like'' DE.
The contours in Fig. \ref{fig:plc_H0_contours} show a clear degeneracy between $a_c$($\alpha$) and $H_0$ in TDE  model. The $\chi^2$ statistic in \ref{tab:plc_H0} shows that, the goodness of the fit is significantly improved. Noted that, this improvement is mainly due to the reduction of the discordance with SH0ES data.

\subsection{Constraints from Accumulative DataSets}
As is discussed in previous sections, the tension between SH0ES and \emph{Planck} is resolved by calibration of the 
transitional scale $a_c$. The problem is now transformed to, whether the expansion history (growth history) fixed by $M^{\rm{R20}}_{\rm{B}}$+\emph{Planck} is in accord with other observations. If the model does restore the cosmic concordance, one would expect consistency between the fit
of any dataset combinations. To verify this, we consider the constraints from accumulative datasets.
We can see from Table \ref{tab:accumulate_constraints}, different combinations of
datasets yield different values of best-fit parameters.
A clear trend should be noticed that, with the inclusion of more datasets, the best-fit cosmological parameters as well as the $\chi^2_{\rm{min}}$ value approach to that of $\Lambda$CDM model. 
For example, with $M^{\rm{R20}}_{\rm{B}}$ and \emph{Planck} combination we find $H_0 = 75.2\pm1.7$ and $S_8 =0.825^{0.013}_{-0.015}$, in $\sim3\sigma$ tension with the fit of the same datasets in $\Lambda$CDM scenario. This tension drops to less than $1\sigma$ with the inclusion of all considered likelihoods, while the differences in $\chi_{min}^2$ value reduce from $23.5$ to $3.5$.
With the inclusion of SNe dataset, the tension with SH0ES increased $1.9\sigma$ for TDE model. Noted that, the SNe 
dataset which constrains the expansion history of the late universe, i.e., $D_{L}(z)/H_0$, is well compatible with 
the $\Lambda$CDM model, thus, modification of the EoS of DE at $a\gtrsim 1/2$ is disfavored.
Consequently, the  $\pm1\sigma$ constrains on transitional scale $a_c$ decrease from $0.658\pm0.049$ to $0.548\pm 0.029$, which results in a decline of the best-fit value of $H_0$.
Interestingly, the tension with SH0ES reduces with the addition of a prior on $S_8 = 0.773^{+0.026}_{-0.020}$ ( $S_8=0.833\pm0.016$ for the \emph{Planck} best-fit $\Lambda$CDM). This is in accord with the shooting result shown in Fig. \ref{fig:shooting_results}, i.e., applying such a prior on $S_8$ would result in a upward shift on the transitional scale $a_c$, which in turn increases the inferred value of $H_0$.
The result of Bayesian evidence analysis shows strong evidence for TDE model when LSS datasets are not considered (see last panel of Table \ref{tab:accumulate_constraints}). While the Bayesian evidence does not show any significant difference between TDE and $\Lambda$CDM model when all datasets are included ($\Delta\ln\varepsilon = 0.15$). 

\subsection{Constraints from Selected DataSets}
In the context of TDE model, the tension with SH0ES increases significantly (from $1.9\sigma$ to $3.0\sigma$) with the inclusion of BAO data. 
As can be seen from Fig. \ref{fig:lss_fs8}, both BAO and SNe measurements (calibrated by the local distance ladder) anchor the distance scale at $z\lesssim1.0$, yet not in good agreement with each other. Such tension is inherited in the dataset itself. Therefore, the conformity with one of the anchor almost certainly leads to discordance with the other.
While within the scenarios of late-DE, the inferred $r_s$ is identical with fiducial $\Lambda$CDM model (see section \ref{sec:ph}), so the discordance between BAO and SNe ($H_0$) datasets can't be reconciled.
However, the BAO measurements rely on the presumption of the fiducial cosmology,
which means the measured distance scale should be calibrated by the factor $r_{\rm{d,fld}}/r_{\rm{d}}$. 
To account for the possibility that the tension is originated by some unknown systematics of a single dataset, we selected several subclasses of whole datasets as follows:

\begin{itemize}
	\item Without BOSS: Comparing the result tabulated in Table \ref{tab:BOSS_SNe} and Table \ref{tab:SH0ES_combined}, we find that 
	with the exclusion of BOSS dataset, 
	the $\pm1\sigma$ constrains on $H_0$ increase from $69.16\pm{0.76}$ to $71.77\pm0.90$ while the tension with SH0ES reduce from $3.2\sigma$ to $1.9\sigma$ (see Fig. \ref{fig:selected_contours} for a view) in TDE model. 
	Which indicates that the tension between BOSS and $M^{\rm{R20}}_{\rm{B}}$ prior derived from SH0ES can not be reconciled in this scenario.  
	
	\item Without SNe: See Table \ref{tab:BOSS_SNe} and Table \ref{tab:SH0ES_combined} for a contrast between the constraint results with and without SNe dataset. 
	As can be seen, the best-fit transitional scale $a_c$ as well as other cosmological parameters (including $H_0$) are almost identical in both case, suggesting that the SNe dataset is well compatible with the TDE scenario.

	\item Without $M^{\rm{R20}}_{\rm{B}}$: From Table \ref{tab:SH0ES_combined} one can see that, the difference 
	in $\pm1\sigma$ constrains on $H_0$ for the two model is marginal, also are the $\chi^2$ statistics, i.e, $\Delta\chi^2=-0.1$. In this case, the extra late-DE parameter is redundant. The numerical results
	suggest that, the transitional scale $a_c$ is consistent with zero, i.e., $a_c<0.252$.
	Meanwhile, the Bayesian evidence is against the TDE model when compared with the fiducial $\Lambda$CDM model ($\Delta \ln\varepsilon =2.74$). 

\end{itemize}
For a brief summary, the TDE scenario is disfavored by the BAO dataset. This discordance could not be fully removed within the context of late-DE. With the exclusion of $M^{\rm{R20}}_{\rm{B}}$ prior and the inclusion of LSS dataset, 
the $\Lambda$CDM model is favored over TDE model.

\section{DISCUSSION AND CONCLUSIONS}\label{sec:conclusion}
The $\Lambda$CDM model calibrated by \emph{Planck} data is well 
compatible with a large bunch of independent observational datasets (BAO, SN Ia, etc), 
but in severe tension with the local distance ladder ($H_0$).
Consequently, the ideal solution to the discordance lies in the scenarios which disagree with 
$\Lambda$CDM solely on the inferences of $H_0$, while recover the predictions of $\Lambda$CDM in most cases.
However, the shift in $H_0$ inevitably leads to the modification in late expansion history. 
A practical way to capture this modification is to add an extra parameter to the six standard $\Lambda$CDM parameters named as ``late-DE parameter'', which degenerates only with $H_0$ while is uncorrelated with the other five parameters.

In this work, we explore a representative parameterization within the scenarios of late dark energy: a novel version of Transitional dark energy (TDE). 
The main feature of the TDE model is the sharp transition of the EoS of DE at the critical scale $a_c$.
When $a_c$ approach to zero, the TDE model recovers $\Lambda$CDM model.
We analyze the TDE scenario accounting for \emph{Planck} 2018 distance prior, BAO+FS (DF6, BOSS DR12 and DR14), $S_8$ prior derived from DES-Y1 3x2pt, as well as the $M^{\rm{R20}}_{\rm{B}}$ prior derived from SH0ES and SNe dataset from the Pantheon compilation.
The results are shown in Table \ref{tab:plc_H0}$\sim$\ref{tab:SH0ES_combined} and Fig. \ref{fig:plc_H0_contours}$\sim$\ref{fig:TDE_MEDE_contours}. To give a clear view on the influences of different datasets, we conduct the MCMC analysis in an accumulative way.  
Due to the degeneracy between $H_0$ and the extra late-DE parameter, $H_0$ is loosely constrained by CMB alone, i.e., $H_0 = 77.0\pm8.0$ for TDE model. When added with the $M^{\rm{R20}}_{\rm{B}}$ prior, the constraint results for TDE ($H_0 = 75.2\pm 1.7$) closely match the SH0ES $H_0$ prior, which further confirms the degeneracy.
With the growing number of included datasets, one can see clear trend that the TDE scenario degenerates into standard $\Lambda$CDM model, meanwhile, the tension with SH0ES grows continuously (see Table \ref{tab:accumulate_constraints}).
Thus, the tension on $H_0$ is not solved, also, the $S_{8}$ tension is neither relieved nor exacerbated in TDE scenario. Owing to the fact that the structure growth is more rapid in the case of ``phantom-like'' DE than that of cosmological constant, a lower inferred value of $\Omega_{m,0}$ is largely compensated by the upward shift on $\sigma_{8}$, resulting in a minor influence on the $S_8$ constraints.

The three main anchors of the cosmic distance scale, e.g., CMB ($z\sim1100$), BAO($z\sim0.3$), and SNe Ia calibrated by the local $H_0$ measurements ($z\lesssim 1.5$) form a ``Impossible trinity'', i.e., it's plausible to reconcile with any two of the three but hard to accommodate all of them.
Within the scenarios of late-DE, it's nearly impossible to reconcile the tension between BAO ($z\sim0.3$) and SNe Ia calibrated by $H_0$ (see Fig. \ref{fig:lss_fs8}). Even if we allow for a reduction of the sound horizon $r_s$, it is not likely to simultaneously match the distance anchored by CMB and BAO (see \citet{DAmico:2020ods,Hill:2020osr}), because given a certain shift in $r_s$, the shift in $H_0$
needed to match the CMB observations will be different than the one needed to match
the late LSS observations. 
However, the late universe measurements require the modeling of complex astrophysical systems which may bring about some systematic
error either in the measurements or in the astrophysical modeling (\citet{DAmico:2020ods, Mortsell:2021nzg}).
To account for the possibility of unknown systematics, we selected several subclasses of the datasets,
the constraint results can be seen in Table \ref{tab:BOSS_SNe}, \ref{tab:SH0ES_combined} and Fig. \ref{fig:selected_contours}.
In the analysis without BOSS measurement, TDE model have shown its potential to solve the $H_0$ tension, i.e., $H_0=71.77\pm0.90$ and $a_c = 0.531^{+0.045}_{-0.040}$ (reconcile with SH0ES within $2\sigma$). This result can also be verified by ``shooting'',
which suggests that, at $a_c=0.531$ the $\emph{Planck}$ best-fit value is $H_0=72.19$ (see Fig. \ref{fig:shooting_results}).
With the combination of all considered datasets (including $M^{\rm{R20}}_{\rm{B}}$), we see a positive sign that the EoS transits in TDE scenario, i.e., the transitional scale of DE is not consistent with zero ($a_c = 0.375^{+0.12}_{-0.043} $ at $68\%$ CL). With the exclusion of th $M^{\rm{R20}}_{\rm{B}}$ prior, we have $a_c$ peaked toward zero. In this case, the result of Bayesian evidence analysis is against TDE model when compared with $\Lambda$CDM model. However, the analysis without BOSS measurement indicates strong evidence for TDE model ($\Delta\ln\varepsilon = 4.06$). 

In summary, owing to the irreconcilable tension between BAO and SNe Ia calibrated by $H_0$, within the scenario of TDE the cosmological concordance can not be restored. 
However, with the presumption of unknown systematics, the TDE model could still be viewed as a viable candidate.
What is the ``true'' origin of these discordances?  We hope that more investigators will be motivated to explore this sector. 

\section*{Acknowledgements}

This work is supported in part by National Natural Science Foundation of China under Grant No.12075042, Grant No.11675032 (People's Republic of China).

\section*{Data Availability}
The data underlying this article will be shared on reasonable request to the corresponding author.



\bibliographystyle{mnras}
\bibliography{TDE} 




\appendix

\section{Prior Dependence}
\begin{figure}
	\centering
	\includegraphics[scale=0.33]{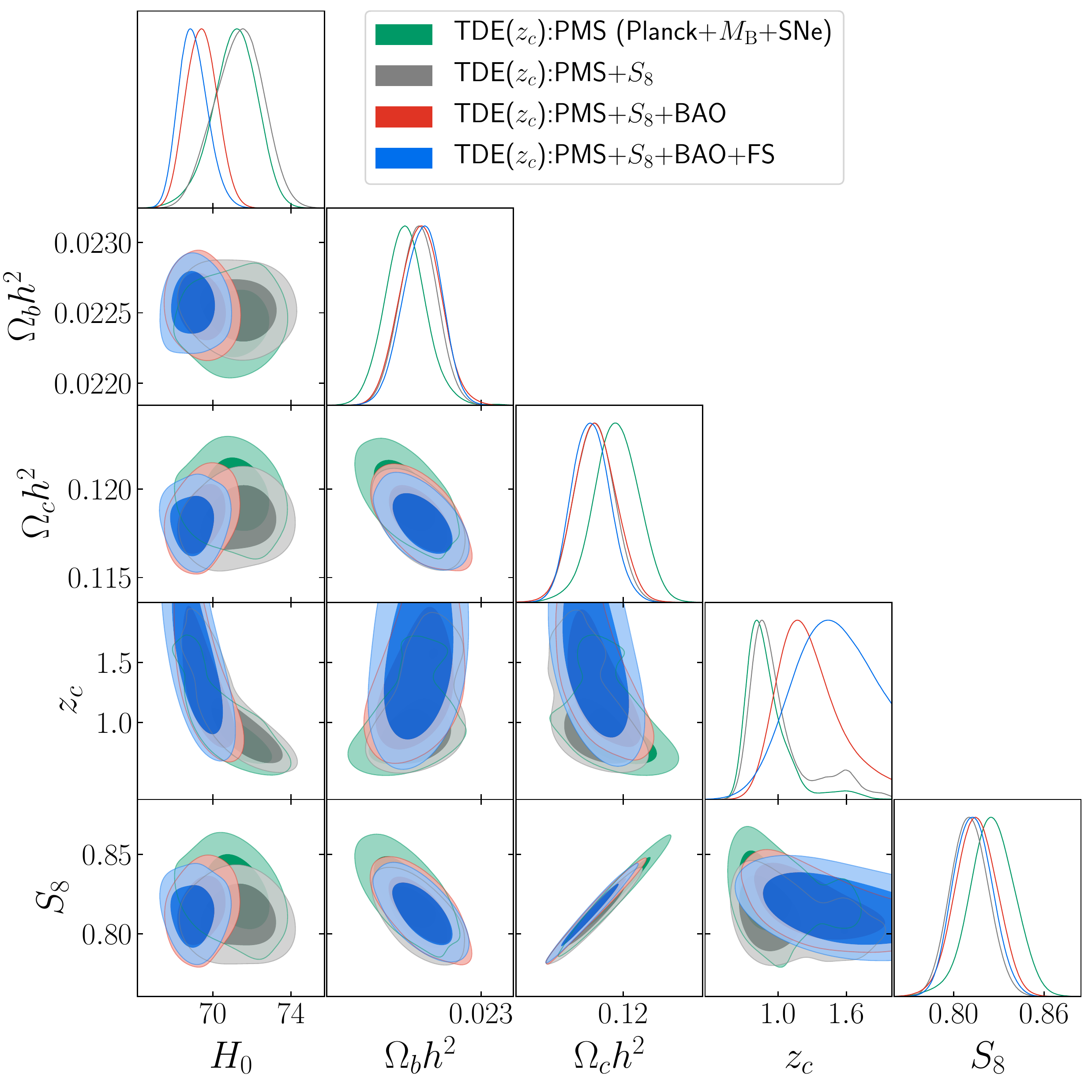}
	\caption{Constraints on TDE scenario with a uniform prior on $z_c$ (denote as TDE($z_c$)) from various dataset combinations.  The contours show $1\sigma$ and $2\sigma$ posteriors. 
		The constraints on cosmological parameters closely match the results shown in the left panel of Fig. \ref{fig:TDE_MEDE_contours}. }
	\label{fig:TDE_zc_countour}
\end{figure}

We have assumed uniform prior probability distributions for effective TDE parameter $a_c$,
which correspond to non-uniform priors on the transitional redshifts $z_c = 1/a_c-1$.
The EoS in parameterized by $z_c$ (denote as TDE($z_c$)) can be rewritten as:
\begin{eqnarray}
w(z) = -1 -\frac{1}{2}\left[\tanh(3(z-z_c))+1\right]
\end{eqnarray}
An obvious concern is the dependence of the posterior distributions on the choice of priors.
To account for that, we recompute the TDE parameter constraints with a uniform prior imposed on $z_c$, i.e., $z_c\in[0,100]$, which 
corresponds to $a_c\in[0.01,1]$. 
The posterior distributions are shown in
Fig. \ref{fig:TDE_zc_countour} and the parameter constraints are tabulated in Table \ref{tab:TDE_TDE2}.
It's obvious to see that, the difference in priors have a negligible impact on the $\chi^2$ statistics, e.g., $\chi^2_{min} = 1068.4$ for the TDE fit to the all combined datasets and  $\chi^2_{min} = 1068.7$ for that of TDE($z_c$) 
fit of all combined datasets. It is notable that, there is a slight decline on the inferred value of $H_0$ in TDE($z_c$), which could be understood as the effect of a shift in prior probability distribution toward a smaller value of $a_c$ when a uniform prior on $z_c$ is assumed. 
The difference between the fit of TDE and TDE($z_c$) becomes insignificant with the inclusion of a growing number of datasets (see Table \ref{tab:TDE_TDE2}).

\small{
	\begin{table*}
		\renewcommand{\arraystretch}{1.2}
		\begin{center}
			\begin{tabular}{|lllllllll|}			
				\hline
				\hline			
				Model\hspace{2pt} &Dataset & $H_0$ & $S_8$ & $a_c$ &   &$\ln\varepsilon$ &$\Delta\ln\varepsilon$&$\chi^2_{\rm{min}}$\\
				\hline
				\hline
				
				TDE& \emph{Planck} + $\rm{M_B}$  & $75.2\pm 1.7                $ & $0.825^{+0.013}_{-0.015}           $ & $0.658\pm 0.049    $& $-$  & -11.54&$-$&$4.6             $\\
				
				&PMS   &  $71.59\pm 0.81             $ & $0.825\pm 0.011            $ &$0.548\pm 0.029               $& $-$ &-533.04&$-$&$1045.0 $\\
				
				&PMS+$S_8$   &  $71.77\pm 0.90     $ & $0.812\pm 0.012    $ & $0.531^{+0.045}_{-0.040}    $ &$-$
				&-535.07&$-$&1051.3\\
				
				&PMS+$S_8$+BAO & $69.55^{+0.73}_{-0.94}            $&  $0.816\pm 0.015            $ & $0.420\pm 0.082     $ & $-$&-540.25&$-$&1062.3\\
				
				&PMS+$S_8$ +BAO +FS &  $69.16\pm 0.76              $ & $0.814\pm 0.012         $ &$0.375^{+0.12}_{-0.043}    $ &$-$&-543.41&$-$&1068.4\\
				\hline
				& & $H_0$ & $S_8$ & $a_c$ & $w_0$ &$\ln\varepsilon$ &$\Delta\ln\varepsilon$ & $\chi^2_{\rm{min}}$\\
				\hline
				TDE(2p)& \emph{Planck} + $\rm{M_B}$   & $74.6\pm 1.5               $ & $0.824\pm 0.014            $ & $0.50^{+0.027}_{-0.033}   $& 
				$> -1.23        $ & -7.52&-4.02 &$4.1             $\\
				&PMS &  $71.44\pm 0.93$ & $0.828\pm0.013$ & $0.638^{+0.078}_{-0.062}  $ &$-0.86^{+0.12}_{-0.14} $ 
				&-532.95&-0.09&$1045.3         $\\
				
				&PMS+$S_8$ & $70.09\pm 0.78  $ & $0.812\pm 0.012$ & $0.652\pm 0.078 $ & $-0.83^{+0.12}_{-0.14}   $
				&-535.17&0.10&$1049.5          $\\
				
				&PMS+$S_8$+BAO & $70.25\pm 0.79 $ & $0.819\pm 0.013 $ & $0.378^{+0.16}_{-0.064} $ &$-1.04^{+0.04}_{-0.05}  $&-539.71&-0.54&$1061.7         $\\
				
				&PMS+$S_8$+BAO+FS & $69.69\pm 0.79 $ & $0.817\pm 0.012 $ & $0.32^{+0.17}_{-0.20} $ &$-1.02^{+0.02}_{-0.07}   $&-543.67&0.26&$1069.0          $\\
				\hline
				& & $H_0$ & $S_8$ & $z_c$ &&&&$\chi^2_{\rm{min}}$\\
				\hline
				TDE($z_c$) & \emph{Planck} +  $\rm{M_B}$   &  $73.8\pm 2.1 $ & $0.825\pm0.014$ & $0.63\pm0.15$ &$-$& 
				$-$&$-$&$-$ \\
				
				&PMS   &  $71.1^{+1.2}_{-1.0} $ & $0.825^{+0.015}_{-0.018}$ & $0.91^{+0.08}_{-0.19}$ &$-$
				&$-$&$-$&$1046.8         $\\
				
				&PMS+$S_8$ & $71.3^{+1.3}_{-1.2}    $ & $0.811\pm 0.012 $ & $1.014^{+0.062}_{-0.30} $ &$-$
				&$-$&$-$&$1051.1         $ \\
				
				&PMS+$S_8$+BAO & $69.43\pm 0.78  $ & $0.815\pm 0.014$ & $1.643^{+0.048}_{-0.79} $ &$-$
				&$-$&$-$&$1061.8          $ \\
				
				&PMS+$S_8$+BAO +FS & $69.0^{+0.66}_{-0.81}$ & $0.813\pm 0.013 $ & $1.88^{+0.38}_{-0.92} $ &$-$
				&$-$&$-$&$1068.7          $\\
				\hline
								& & $H_0$ & $S_8$ & $a_c$ & $\tau_{\rm{reio}}$& $\log(10^{10}A_s)$&$n_s$&$\chi^2_{\rm{min}}$ \\
				\hline
				TDE & \emph{Planck}:TTTEEE +  $\rm{M_B}$   &  $74.8\pm 1.9 $ & $0.812^{+0.018}_{-0.015} $ & $0.613^{+0.054}_{-0.047}$ &$0.0527^{+0.0087}_{-0.0074}$& 
				$3.039^{+0.020}_{-0.014}$&$0.9665^{+0.0042}_{-0.0060}$&$1012 $ \\
				
				&TMS   &  $72.0\pm 1.2 $ & $0.818\pm 0.017  $ & $0.535^{+0.042}_{-0.034}$ &$0.0522^{0.0085}_{-0.0086}$&
				$3.039^{0.017}_{-0.018}        $&$0.9660^{0.0041}_{-0.0043}$&$2050$\\
				
				&TMS+$S_8$ & $71.9^{+1.2}_{-0.90}    $ & $0.807\pm 0.014 $ & $0.517^{+0.041}_{-0.032} $ &$0.0516^{+0.0088}_{-0.0077}$& $3.035^{+0.020}_{-0.014}         $&$0.9673^{+0.0054}_{-0.0037}$&$2057$ \\
				
				&TMS+$S_8$+BAO & $69.84^{+0.64}_{-0.84}  $ & $0.812^{+0.015}_{-0.011}$ & $0.398^{+0.10}_{-0.031} $ &$0.0536^{+0.0075}_{-0.0096}$&$3.039^{0.017}_{-0.018}          $&$0.9685^{+0.0042}_{-0.0054} $&$2068$ \\
				
				&TMS+$S_8$+BAO +FS & $69.29^{+0.67}_{-0.77}$ & $0.811\pm 0.013 $ & $0.351^{+0.12}_{-0.045} $ &$0.0530^{+0.0072}_{-0.0094}$&$3.037^{+0.015}_{-0.019}      $&$0.9682^{0.0045}_{-0.0047}   $&$2072$\\
				\hline

			\end{tabular}
		\end{center}
		\caption{The the mean $\pm1\sigma$ constraints on the cosmological 
			parameters in TDE model parameterized by $a_c$ (first panel), TDE(2p) parameterized by $a_c$ and $w_0$ (second panel), TDE model parameterized by $z_c$ (third panel), and TDE model constrained by \emph{Planck} TT+TE+EE data and
			several dataset combinations (fourth panel). Noted that the pair difference of Bayesian evidence is given by: $\Delta\ln\varepsilon = \ln\varepsilon_{\rm{TDE}}-  \ln\varepsilon_{\rm{TDE(2p)}}$.}
		\label{tab:TDE_TDE2}
	\end{table*}
}

\section{Two Parameters TDE(2p)}\label{sec:TDE2pars}
\begin{figure}
	\centering
	\includegraphics[scale=0.28]{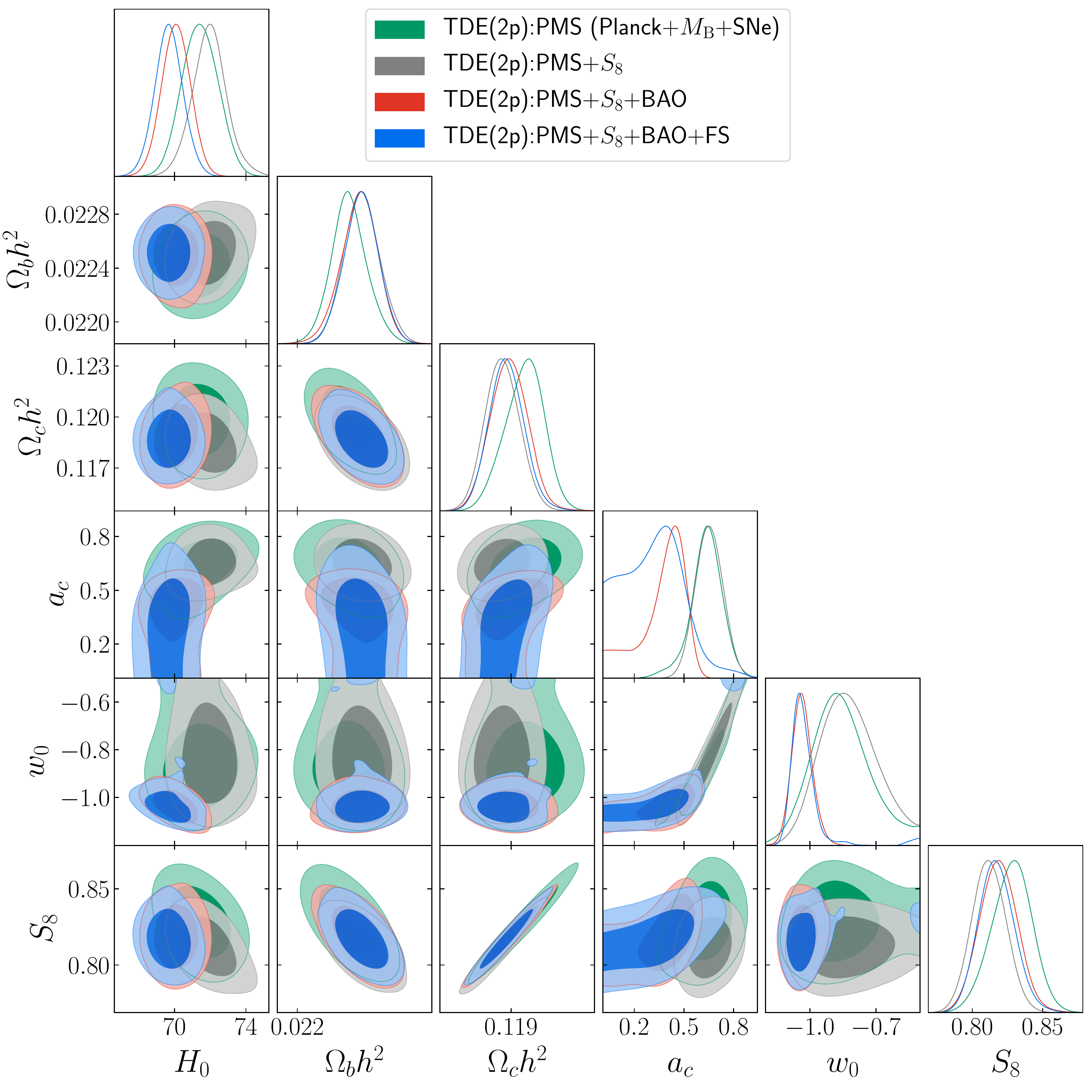}
	\caption{Constraints on two parameters TDE(2p) model from various dataset combinations.  The contours show $1\sigma$ and $2\sigma$ posteriors. See Table \ref{tab:TDE_TDE2} for the mean $\pm 1\sigma$ constraints on the cosmological parameters.}
	\label{fig:TDE_2p_countour}
\end{figure}

In previous sections we have fixed $w_0 = -1$ to recover $\Lambda$CDM when $a\gtrsim a_c$.
While in this section,  $w_0$ is considered as a free constant. In this case, the transitional DE is described 
by two parameters ($w_0$, $a_c$).
The parameter constrains and the posterior distributions are 
shown in  Table \ref{tab:TDE_TDE2} and Fig. \ref{fig:TDE_2p_countour}.
From the middle panel of Table \ref{tab:TDE_TDE2} one can see that, the constrains in TDE(2p) model 
on $\Omega_bh^2, \Omega_ch^2$ and $S_8$ from each dateset combination closely match that of TDE(1p) model,   
while the constraint on $w_0$ is in agreement with -1 within $2\sigma$.
As can be noticed, there is a slight downward shift on the fit of $w_0$ from $w_0\gtrsim-1$ to $w_0\lesssim -1$ with the inclusion of BAO datasets. As is discussed in section \ref{sec:models}, the downward shift on $w_0$ would increase the dilution rate of DE, which lowers inferred value of $\theta_{\rm{LSS}}$ while 
increases the growth function $f(z)$ in the late universe.
As to $\chi^2$ statistics, adding a free parameter ($w_0$) has negligible improvement on the goodness of the fit 
(see the third panel of Table \ref{tab:TDE_TDE2}).
So effectively speaking, the TDE model is efficiently captured by a single parameter $a_c$.
As is shown in the last column of Table \ref{tab:TDE_TDE2}, we have compared the Bayesian evidence between TDE and TDE(2p) model. In the case of all datasets combination, we find $\Delta\varepsilon = 0.26$ indicating the weak evidence against TDE(2p).

\section{\emph{Planck} TTTEEE Results}
\label{sec:planck}
\begin{figure*}
	\centering
	\includegraphics[scale=0.36]{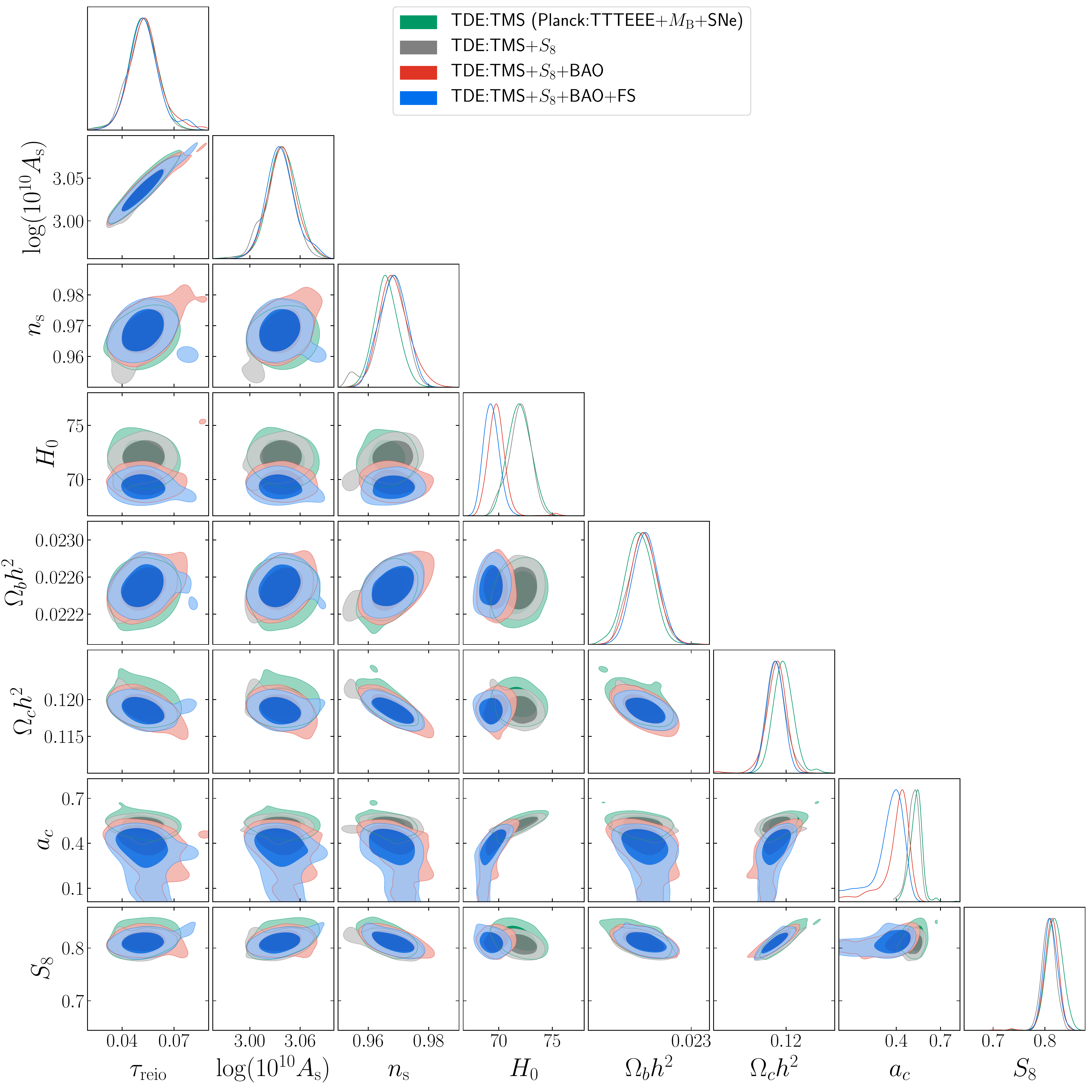}
	\caption{Constraints on TDE scenario with \emph{Planck} TT+TE+EE data and several dataset combinations.  The contours show $1\sigma$ and $2\sigma$ posteriors.}
	\label{fig:TDE_TTTEEE_countour}
\end{figure*}

\minew{The constraint results from the \emph{Planck} 2018 low-$\ell$ TT+EE and \emph{Planck} 2018 high-$\ell$ TT+TE+EE temperature and polarization power spectrum and several dataset combinations 
are tabulated in the last panel of Table \ref{tab:TDE_TDE2},  and the posterior
distributions are shown in Fig. \ref{fig:TDE_TTTEEE_countour}. Note that, the constraints on $H_0$, $a_c$ and $S_8$ from \emph{Planck} TT+TE+EE data closely
matches the results obtained from the \emph{Planck} distance prior.}

\bsp	
\label{lastpage}
\end{document}